\documentclass[utf8]{aa}  
\usepackage{orcidlink}

\usepackage{graphicx}
\usepackage{txfonts}
\usepackage{hyperref}
\usepackage{siunitx}
\DeclareSIUnit\year{yr}
\DeclareSIUnit\erg{erg}
\DeclareSIUnit\pc{pc}
\usepackage{pgfplots}
\pgfplotsset{compat=1.14}
\usepackage{layouts}
\usepackage[switch]{lineno}
\modulolinenumbers[5]

\begin{document}

\title{A novel pathway to stellar-mass black hole growth}

\subtitle{Re-accretion of gas from stellar wind bubbles}

\author{D. Kom\'anek
      \inst{1,2}
      \orcidlink{0009-0004-1164-2604}
      \and
      R. W\"unsch\inst{1}
      \orcidlink{0000-0003-1848-8967}
      \and
      S. Mart\'{\i}nez-Gonz\'alez\inst{3}
      \orcidlink{0000-0002-4371-3823}
      \and
      S. Jim\'enez\inst{1}
      \orcidlink{0000-0003-2808-3146}
      \and
      D. Sz\'ecsi\inst{4}
      \orcidlink{0000-0001-6473-7085}
      }

\institute{Astronomical Institute of the Czech Academy of Sciences,  Bo\v{c}n\'\i\ II 1401/1, 141 00 Praha 4, Czech Republic\\
              \email{komanek@asu.cas.cz}
         \and
             Faculty of Mathematics and Physics, Charles University, V Hole\v{s}ovi\v{c}k\'ach 2, 180 00 Praha 8, Czech Republic
         \and
             Instituto Nacional de Astrof\'\i sica \'Optica y Electr\'onica, AP 51, 72000 Puebla, M\'exico
         \and
             Institute of Astronomy – Faculty of Physics, Astronomy and Informatics – Nicolaus Copernicus University, Grudzi\k{a}dzka
    5, 87-100 Toru\'n, Poland
            }


  \abstract
   {The formation of high-mass stellar black holes (BHs) observed in mergers by gravitational wave observatories remains difficult to explain using standard models of stellar evolution.}
   {We investigate whether newly formed stellar-mass BHs in dense molecular clouds can significantly increase their final mass by accreting gas from pre-existing, wind-blown interstellar bubbles produced by their progenitor stars.}
   {We performed 1D and 3D hydrodynamic simulations of interstellar bubbles driven by stellar feedback, incorporating thermal conduction, radiative cooling, gravity, ionizing radiation, stellar winds, and supernova (SN) explosions. We modeled the subsequent accretion onto the central BH for two formation scenarios: core-collapse SN and direct collapse. Our parameter space study encompasses over 500 1D simulations, varying initial cloud densities, gravitational binding energies, progenitor masses ($40-257\, M_\odot$), and metallicities (\(0.66\,Z_\odot\) and \(0.016\,Z_\odot\)), and tracks the bubble evolution for millions of years post-BH formation. Scenarios exhibiting significant re-accretion were further validated using high-resolution 3D simulations.}
   {Our simulations show that BHs formed via direct collapse from low-metallicity progenitors can accrete up to \SI{6.9}{M_\odot} of gas from massive inner shells formed by late-stage slow winds, resulting in final BH masses of up to \SI{93}{M_\odot}. Conversely, we find no significant re-accretion in scenarios involving an SN explosion or high-metallicity progenitors; in these cases, fast winds or SN ejecta efficiently evacuate dense structures from the bubble interior, leaving behind a hot, low-density medium.}
   {We demonstrate that while SN explosions and fast stellar winds suppress post-formation accretion by redistributing mass into cold outer shells, direct collapse in low-metallicity environments provides a viable pathway for BHs to gain substantial mass from their circumstellar surroundings. This mechanism may help explain the unexpectedly high masses of BHs observed in gravitational wave events.}


   \maketitle
%
\section{Introduction}
\label{sec:introduction}

Recent observations made by the Advanced LIGO \citep{aasi_advanced_2015}, Advanced Virgo \citep{acernese_advanced_2015}, and KAGRA \citep{abbott_prospects_2020} gravitational wave (GW) observatories have revealed a population of stellar black holes (BHs) with masses up to \SI{137}{M_\odot} \citep{abac_gwtc-40_2025}. This finding far exceeded previous expectations and presents a serious challenge to models of stellar-mass BH formation. This high-mass GW population stands in stark contrast to the historically established mass range of stellar-mass BHs, which are significantly lighter. Observations of stellar-mass BHs via X-ray binaries, astrometry, and microlensing have predominantly yielded masses below \SI{20}{M_\odot} \citep{casares_mass_2014, ramachandran_comprehensive_2025, el-badry_sun-like_2023, panuzzo_discovery_2024, sahu_isolated_2022}.

There are several distinct pathways through which stellar-mass BHs can form: the core-collapse supernova (SN) of a~massive star, the direct collapse of a stellar core without a visible explosion, and the merger of compact objects, such as two neutron stars or lower-mass BHs \citep{mirabel_formation_2017, gerosa_hierarchical_2021}. The properties of the resulting stellar-mass BHs depend heavily on the evolution of single massive stars, stars in multiple stellar systems, and stars in dense environments such as globular clusters and gaseous disks of active galactic nuclei (AGN). These processes introduce significant theoretical uncertainties, making it difficult to draw confident conclusions about their formation.

Focusing on the isolated single-star pathway, the specific outcome for a massive star is heavily dependent on its initial mass and metallicity \citep{heger_how_2003}. While some undergo core-collapse or direct collapse to form BHs, others may explode as pair-instability SNe, leaving no compact remnant. The upper mass of a BH formed from a single star is limited by the star's initial mass. While most massive stars are born with initial masses up to \SI{150}{M_\odot} \citep{figer_upper_2005, koen_upper_2006, martins_most_2008}, limited observational evidence suggests that stars with initial masses up to \SI{300}{M_\odot} may exist in the local universe \citep{crowther_r136_2010}. The mass budget available for BH formation is further reduced by mass loss through stellar winds and SN ejecta. More metal-rich stars experience higher rates of mass loss through stellar winds \citep{heger_how_2003, vink_metallicity-dependent_2021}. However, not all ejecta necessarily escape; a~fraction of the stellar envelope may remain bound or be decelerated by reverse shocks, eventually accreting back onto the central compact remnant. This process, known as fallback accretion, is a~primary determinant of the final compact remnant mass \citep{chevalier_neutron_1989, zhang_fallback_2008, fryer_compact_2012}. While fallback typically occurs on short timescales following the explosion, this study explores whether the BH can accrete mass from the wider environment over much longer timescales.

Several studies have explored which massive stars explode as SNe and what the masses of the resulting compact remnants will be \citep{ugliano_progenitor-explosion_2012, pejcha_landscape_2015, ertl_two-parameter_2016, sukhbold_core-collapse_2016, muller_simple_2016, ebinger_pushing_2019}. They show that there is no distinct threshold that divides the mass spectrum of stars into those that explode and those that directly collapse into a BH. Instead, there are regions where stars experience an explosion among those that undergo direct collapse. These regions are called islands of explodability. Because there is no simple mass threshold dictating which stars explode and which silently collapse, it is crucial to investigate how a pre-existing wind-blown bubble responds to both scenarios to fully understand the re-accretion potential.

The cores of massive stars that are sufficiently hot (\(> \SI{e9}{\kelvin}\)) but not too dense can undergo a partial or complete collapse due to the creation of electron-positron pairs---a process called pair instability \citep{fowler_neutrino_1964}. In helium cores with masses greater than \SI{133}{M_\odot}, nuclear burning cannot stop the collapse, and a BH forms \citep{fryer_pair-instability_2001, heger_nucleosynthetic_2002}. Helium cores with masses above \SIrange[range-phrase = \text{--}]{60}{65}{M_\odot} explode as a pair-instability SN---a single pulse that completely disrupts the star, leaving no compact remnant. Lighter helium cores with masses above \SIrange[range-phrase = \text{--}]{40}{45}{M_\odot} experience multiple pulses that are too weak to completely unbind the star. Each time, the star loses mass. This process continues until the core’s mass falls below the \SIrange[range-phrase = \text{--}]{40}{45}{M_\odot} threshold. The pulsationally-ejected layers may get shocked and give rise to a so-called pulsational pair-instability SN. Lighter stars may form either a~neutron star or a~BH, depending on the mass loss they undergo throughout their evolution.

The mass ejection or complete destruction of stars due to pair instability should result in an approximately \(\SIrange[range-phrase = \text{--}]{50}{133}{M_\odot}\) mass gap in stellar-mass BHs formed from single stars \citep{stevenson_impact_2019}. Depending on the assumptions used, some models shift the lower boundary, resulting in a narrower \(\SIrange[range-phrase = \text{--}]{70}{133}{M_\odot}\) mass gap \citep{belczynski_effect_2016}. However, BHs within this mass gap have been observed via GWs \citep{abac_gwtc-40_2025}, as mentioned above. Several mechanisms have been proposed to populate the mass gap, including uncertainties in stellar evolution models \citep{belczynski_effect_2016, stevenson_impact_2019}, mass transfer in close binaries \citep{van_son_polluting_2020}, and hierarchical mergers \citep{gerosa_hierarchical_2021}. However, these processes are still awaiting strong observational confirmation, as explained below. Therefore, we study another scenario.

Regarding the single-star pathway, the exact boundaries of the pair-instability mass gap are sensitive to uncertainties in stellar evolution, such as mixing, rotation, and mass-loss rates \citep{ekstrom_grids_2012, spera_very_2017, stevenson_impact_2019}. For instance, if a star manages to retain its massive hydrogen envelope after a pulsation, it can theoretically leave behind a BH of up to \SI{70}{M_\odot} \citep{belczynski_effect_2016}. However, pushing single-star models to routinely produce remnants near \SI{100}{M_\odot} without invoking extreme or fine-tuned physics remains a severe challenge \citep{woosley_pulsational_2017, belczynski_evolutionary_2020}.

The second proposed pathway involves binary interactions. Most massive stars are found in multiple stellar systems \citep{duchene_stellar_2013, sota_galactic_2014}, where it may be possible for a BH to accrete mass from a stellar companion. For this to occur, the BH must not receive a natal kick strong enough to eject it from the system. To explain the mergers observed through GWs, models must not only predict the formation of massive BHs, but also their subsequent merging. \citet{van_son_polluting_2020} used a population synthesis code to study the evolution of isolated binaries. They found that, although binaries with a~total mass within the pair instability mass gap should exist, the components are typically too far apart to merge within a Hubble time.

Finally, stellar-mass BHs may also form through hierarchical mergers of multiple BHs in dense environments, such as globular clusters \citep{portegies_zwart_runaway_2002, gultekin_three-body_2006}, or AGN disks \citep{mckernan_intermediate_2012, bartos_rapid_2017, yang_hierarchical_2019}. However, mergers of spinning BHs produce asymmetric GW emission, giving the merged BHs kicks on the order of thousands of \si{\kilo\meter\per\second}, preventing subsequent mergers \citep{gonzalez_supermassive_2007, campanelli_maximum_2007}. Environments with high escape velocities are needed if BHs form with significant spin \citep{antonini_merging_2016}.

While dynamical and binary interactions offer potential pathways to bridge the mass gap, the immediate circumstellar environment of single massive stars provides another, largely unexplored mass reservoir. Massive stars shape their immediate surroundings through fast stellar winds and ionizing UV radiation, inflating hot, pressurized bubbles bounded by a dense shell of swept-up interstellar medium. The structure and evolution of these bubbles have been described in many theoretical works \citep{avedisova_formation_1972, castor_interstellar_1975, weaver_interstellar_1977, koo_dynamics_1992, toala_radiation-hydrodynamic_2011, geen_detailed_2015, el-badry_evolution_2019}. A typical stellar bubble is composed of a~free wind region consisting of wind that reaches velocities of up to thousands of \si{\kilo\meter\per\second}. At the reverse shock, most of the kinetic energy of the gas transforms into thermal energy as it enters the shocked wind region. The hot gas, with temperatures in the millions or tens of millions of kelvins, expands and collects the ambient medium at the edge of the bubble, forming a~thin, dense shell of swept-up interstellar medium. However, the ambient medium can be ionized by UV radiation from the star. Therefore, the shocked wind region may be surrounded by an \ion{H}{ii} region. A~contact discontinuity separates the shocked wind region from the \ion{H}{ii} region. An ionization front then separates the \ion{H}{ii} region from the shell. Finally, a shock front separates the shell from the ambient medium.

\citet{tenorio-tagle_evolution_1990, dwarkadas_evolution_2005} studied the case of an SN exploding inside an interstellar bubble using 1D spherically symmetric models. They demonstrated that, when the ejecta reaches an interface with a~changing density, two waves are created: a~transmitted wave and a~reflected wave. This process can repeat multiple times, resulting in an intricate pattern of reverberating transmitted and reflected waves. Consequently, an SN explosion can result in complex mass redistribution inside a bubble. Subsequent 2D simulations by \citet{franco_evolution_1991, tenorio-tagle_evolution_1991, dwarkadas_evolution_2007} confirmed that the global expansion dynamics remain similar to the 1D cases. However, they found that hydrodynamic instabilities and turbulence inside the bubble cause mixing that washes out multiple reverberations seen in spherically symmetric models. Later studies incorporated asymmetries and more complex geometries. \citet{rozyczka_evolution_1993, brighenti_evolution_1994} examined the effects of moving progenitors and off-center explosions, while \citet{meyer_asymmetric_2015, meyer_wind_2020} focused on runaway stars and SN remnants interacting with bow shocks. Detailed studies of feedback efficiency and pre-processing of the environment have been conducted using high-resolution 1D models \citep{haid_supernova_2016, fierlinger_stellar_2016}, while full 3D radiation-hydrodynamic simulations have explored the complex morphology of these interactions \citep{geen_detailed_2015, martinez-gonzalez_supernovae_2019}.

Observational studies often attribute peculiar morphologies or emission properties to SNe exploding within cavities. Several remnants show strong evidence of expanding inside pre-existing wind-blown bubbles \citep[e.g.,][]{levenson_rosat_1997, borkowski_circumstellar_1996, bauer_supernova_2008}.
In contrast, X-ray studies indicate that most Type Ia SNe evolve in a homogeneous medium unperturbed by the progenitor, unlike their core-collapse counterparts \citep{badenes_are_2007}. The Kepler SN remnant represents a rare Type Ia exception, showing clear signs of ejecta interacting with a dense bow-shock or shell structure \citep{borkowski_dynamics_1992}.

Standard models of compact object formation assume that material launched from a dying star with positive total energy escapes the system permanently, except for a small fraction that undergoes early fallback. In this paper, we explore an alternative mechanism. A massive star embedded in a dense cloud inflates a~wind-driven bubble surrounded by a~dense, cold shell. When the star ends its life (either by exploding as a core-collapse SN or by directly collapsing into a~BH), the resulting dynamics inside this pre-existing cavity dictate the ultimate fate of the circumstellar gas. We systematically investigate the hydrodynamical evolution of these systems to determine how the interplay between the explosion (or lack thereof), stellar wind history, radiative cooling, and environmental density shapes the bubble's interior. Specifically, we aim to map the conditions under which the outward expansion of gas is halted, assessing whether any evolutionary pathways allow material to eventually re-accrete onto the central BH. This process, if efficient, would effectively raise the upper mass limit for BHs formed from isolated single stars.

The paper is structured as follows: In Section \ref{sec:model}, we describe the physical and numerical models used in our 1D and 3D simulations. In Section \ref{sec:results}, we present the results of our comprehensive 1D parameter space study and subsequently examine the most critical scenarios in 3D. In Section \ref{sec:discussion}, we discuss the astrophysical implications of our findings. We summarize our conclusions in Section \ref{sec:conclusions}.

\section{Model}
\label{sec:model}

The model consists of a homogeneous, spherically symmetrical cloud containing a star at its center. The star interacts with the cloud through its stellar wind, ionizing radiation, and gravity. Ultimately, the star either explodes as an SN or undergoes a~direct collapse. In both cases, a stellar-mass BH is left behind. We measure how much gas from the cloud, stellar wind, and SN ejecta can be accreted by the BH over time.

\subsection{Hydrodynamics and gravity}
\label{sec:hydro}

We use the three-dimensional adaptive mesh refinement code \textsc{Flash} 4.6.2 \citep{fryxell_flash_2000}. The hydrodynamic equations are solved using the directionally split piecewise parabolic method \citep{colella_piecewise_1984} advanced via Strang splitting. To compute the time-advanced fluxes at cell boundaries, the code uses a two-shock approximate Riemann solver extended for general equations of state \citep{1985JCoPh..59..264C}. We enable contact steepening, disable flattening, and apply a von Neumann-Richtmyer artificial viscosity coefficient of \(0.1\). To guarantee mass conservation, we use 0th order interpolation between different grid refinement levels. Using a~time step governed by the Courant--Friedrichs--Lewy condition \citep{courant_uber_1928}, \textsc{Flash} solves the following equations of compressible flow
\begin{equation}
    \frac{\partial \rho}{\partial t} + \nabla \cdot \rho \mathbf{v} = q_\mathrm{m}\,,
    \label{eq:f1}
\end{equation}
\begin{equation}
    \frac{\partial \rho \mathbf{v}}{\partial t} + \nabla \cdot \rho \mathbf{v}\mathbf{v} = \rho \mathbf{g} - \nabla P - q_\mathrm{m} \mathbf{v} + \mathbf{q}_\mathrm{p}\,,
    \label{eq:f2}
\end{equation}
\begin{equation}
    \frac{\partial \rho E}{\partial t} + \nabla \cdot (\rho E + P) \mathbf{v} = \rho \mathbf{v} \cdot \mathbf{g} - Q + q_\mathrm{e} - \nabla \cdot \mathbf{q}_\mathrm{c}\,,
    \label{eq:f3}
\end{equation}
where \(t\) is the time; \(\rho\), \(\mathbf{v}\), and \(P\) are the fluid density, velocity, and pressure, respectively; \(\mathbf{g}\) is the acceleration due to gravity, and \(Q\) is the cooling rate. The mass, momentum, and energy deposition rates, denoted as \(q_\mathrm{m}\), \(\mathbf{q}_\mathrm{p}\) and \(q_\mathrm{e}\) respectively, represent the mass, momentum, and energy injected by stellar winds and SNe, and \(\mathbf{q}_\mathrm{c}\) is the heat conduction flux (see Section \ref{sec:conduction}). \(E\) is the sum of internal and kinetic energy per unit mass defined as
\begin{equation}
    E = \frac{P}{(\gamma-1) \rho} + \frac{v^2}{2}\,,
    \label{eq:E}
\end{equation}
where \(\gamma\) is the ratio of specific heats. We set \(\gamma = 5/3\) globally. While a classical diatomic gas has \(\gamma = 7/5\), maintaining \(\gamma = 5/3\) is physically appropriate in our molecular regime because the dense molecular gas quickly cools to \(T \lesssim \SI{100}{\kelvin}\), where temperatures are too low to excite the rotational levels of H\(_2\) \citep[e.g.,][]{2016MNRAS.463.2553R}. The pressure is given by the ideal gas equation of state
\begin{equation}
    P = \frac{k_\mathrm{B} \rho T}{\mu m_\mathrm{H}}\,,
    \label{eq:eos}
\end{equation}
where \(\mu m_\mathrm{H}\) is the mean mass per particle. For the sole purpose of calculating the mean mass per particle in the equation of state, the chemical state of the gas is calculated from the temperature. We assume that gas with a temperature \(T > \SI{9700}{\kelvin}\) is fully ionized, \(\SI{9300}{\kelvin} > T > \SI{700}{\kelvin}\) is atomic, and \(\SI{300}{\kelvin} > T\) is molecular. Linear interpolation is used to calculate \(\mu\) in the transitions between the intervals.

We consider the gravity of stars and compact objects, as well as the self-gravity of gas. We calculate the mass of the BH as the star's mass at the end of its life minus the SN ejecta mass. We update the mass and the gravitational potential as the BH accretes mass. In 1D, spherically symmetric simulations, the self-gravity of the gas is implemented using a simple integrator, as described in \citet{kourniotis_simulations_2023}, which sums the mass enclosed under a~given spherical shell and treats it as a point mass at the center. In 3D simulations, we use the parallel octal-tree based solver developed for \textsc{Flash} by \citet{wunsch_tree-based_2018}.

\subsection{Stellar winds and radiation}

\begin{table}[t!]
    \begin{center}
    \caption{Properties of stars used in the simulations.}
    \label{tab:stars}
    \begin{tabular}{ll}
    \hline
        \(Z_* / Z_\odot\) & \(M_* / M_\odot\) \\ \hline
        \(0.66\) (Milky Way) & \(40, 60, 80, 120, 250\) \\
        \(0.016\) (I Zwicky 18) & \(45, 77, 100, 150, 257\) \\ \hline
    \end{tabular}
    \end{center}

    \vspace{2mm}
    \textbf{Notes.} A list of zero-age main-sequence (ZAMS) masses, \(M_*\), is provided for each respective ZAMS metallicity, \(Z_*\).
\end{table}

We use the Bonn Optimized Stellar Tracks \citep{szecsi_bonn_2022} to follow the evolution of stars with initial masses of \SIrange[range-phrase = \text{--}]{40}{257}{M_\odot}. In Table \ref{tab:stars}, we list the specific zero-age main-sequence masses of the stars used for each metallicity.
The tracks provide the effective temperature, \(T_\mathrm{eff}\), radius, \(R_*\), mass, \(M_*\), and mass-loss rate, \(\dot{m}\), throughout the lifetime of the star. The power of the wind is given by
\begin{equation}
    L_\mathrm{wind} = \frac{1}{2} \dot{m} v_\infty^2\,,
\end{equation}
where \(v_\infty\) is the terminal velocity of the wind. The details of the calculation of terminal velocity can be found in Appendix \ref{app:vinf}.

We implement stellar winds by continuously acting as a source term integrated over each computational time step \(\Delta t\).
The spatial distribution of the injected wind follows the generalized Schuster profile \citep{ninkovic_generalised_1998}. We define a normalized shape function
\begin{equation}
    \rho_\mathrm{S}(r) = N \left ( 1 + \left ( \frac{r}{R_\mathrm{core}} \right )^2 \right )^{-\beta} \qquad \text{for} \quad r < R_\mathrm{cut}\,,
    \label{eq:schuster}
\end{equation}
where \(r\) is the radial coordinate, \(R_\mathrm{core}\) is the core radius, \(\beta\) is a~constant, \(R_\mathrm{cut}\) is the insertion radius, and \(N\) is a normalization constant calculated from the condition \(\int_0^{R_\mathrm{cut}} 4 \pi r^2 \rho_\mathrm{S}(r) \, dr = 1\). We set \(\beta = 1.5\) and \(R_\mathrm{core} = R_\mathrm{cut}/3\). To resolve the wind, we choose to have \(R_\mathrm{cut}\) span 9 cells of the simulation in 1D and 6 cells in 3D.
The mass injection rate per unit volume, corresponding to the source term \(q_\mathrm{m}\) in Equation \ref{eq:f1}, is given by \(q_\mathrm{m}(r)=\dot{m} \rho_\mathrm{S}(r)\). Consequently, the central wind injection rate is \(\dot{m}N\). During each time step \(\Delta t\), the cell density is updated by \(\Delta\rho = q_\mathrm{m} \Delta t\). The treatment of the wind's mechanical energy differs between the 1D and 3D simulations due to spatial resolution constraints. In the high-resolution 1D models, we inject thermal energy at a rate of \(q_\mathrm{e}(r) = 0.5 \dot{m} v_\infty^2 \rho_\mathrm{S}(r)\). Because the injection region is highly resolved, thermal pressure drives the expansion and accelerates the gas to \(v_\infty\) on a scale comparable to the insertion radius, leading to wind formation. In the 3D models, the lower spatial resolution could result in numerical overcooling, radiating the energy away before it could do the mechanical work required to drive the wind. To prevent this, the 3D models inject momentum directly at a rate of \(q_\mathrm{p} (r) = \dot{m} v_\infty \rho_\mathrm{S}(r)\).

We assume that the star radiates as a black body with effective temperature \(T_\mathrm{eff}\). The number of ionizing photons emitted by the star per second can be estimated from Planck's law as
\begin{equation}
    n_\mathrm{UV} = 4 \pi R_*^2 \int_{\nu_\mathrm{min}}^\infty \frac{2 \pi \nu^2}{c^2} \frac{1}{e^\frac{h \nu}{k_\mathrm{B} T_\mathrm{eff}} - 1} \, d\nu\,,
    \label{eq:nuv}
\end{equation}
where \(R_*\) is the radius of the star, \(\nu\) is the frequency of the radiation, \(h\) is Planck's constant, and \(\nu_\mathrm{min} = \SI{13.6}{e \volt} / h \approx \SI{3.3e15}{\hertz}\) is the minimum frequency of ionizing photons. We assume that all the photons created by recombination are immediately reabsorbed (on-the-spot approximation) and that the recombination coefficient to the excited states is \(\alpha_B = \SI{2.6e-13}{\centi\meter^3\second^{-1}}\). In 1D simulations, the radiation transport is solved using a simple integrator, as described in \citet{kourniotis_simulations_2023}. In 3D, we used the TreeRay radiation transport algorithm \citep{wunsch_tree-based_2021}.

\subsection{Supernova injection}

We simulate the SN explosion by adding mass and kinetic energy into a spherical region around the star. Because we cannot resolve the internal explosion processes accelerating the ejecta, neither in space nor time at these evolutionary scales, the injection is modeled as a discrete event at time \(t_\mathrm{SN}\). For the ejecta, we used the density profile from \citet{truelove_evolution_1999} given by
\begin{equation}
  \rho_\mathrm{ej}(r,t) = \frac{M_\mathrm{ej}}{R_\mathrm{ej,cut}^3}f \left (\frac{r}{R_\mathrm{ej,cut}} \right )\,,
  \label{eq:rho_ej}
\end{equation}
where \(R_\mathrm{ej,cut}\) is the insertion radius, \(M_\mathrm{ej}\) is the mass of the ejecta (the specific values of which are detailed in Section \ref{sec:setup}), and \(f\) is a function of the radial coordinate \(r\) defined as
\begin{equation}
  f(r)=\begin{cases}
    f_0 & \qquad \text{for} \quad 0\le r \le R_\mathrm{ej,core}\\
    f_0 (R_\mathrm{ej,core}/r)^n  & \qquad \text{for} \quad R_\mathrm{ej,core}\le r \le R_\mathrm{ej,cut}  \,,
  \end{cases}
  \label{eq:f}
\end{equation}
where \(f_0\) is a constant, \(R_\mathrm{ej,core}\) is the core radius of the profile, and \(n\) is the steepness of the profile. We set \(R_\mathrm{ej,core} = R_\mathrm{ej,cut}/3\) and \(n=2\).\footnote{We tried values up to \(n=11\) without observing any significant changes to the results. We also tried profiles without flat cores (\(R_\mathrm{ej,core} = 0\)) which gave the same results.} To resolve the profile, the insertion radius, \(R_\mathrm{ej,cut}\), is set to span at least 5 cells of the simulation in both 1D and 3D. The constant \(f_0\) is obtained from the normalization condition \(M_\mathrm{ej} = \int_0^{R_\mathrm{ej,cut}} \rho_\mathrm{ej}
4\pi r^2 \,dr\). The velocity profile of the ejecta is given by \citep{truelove_evolution_1999}
\begin{equation}
  v_\mathrm{ej}(r)= v_0 r \qquad \text{for} \quad 0 \le r \le R_\mathrm{ej,cut}\,,
\end{equation}
where the normalization constant \(v_0\) follows from \(E_\mathrm{ej} = \int_0^{R_\mathrm{ej,cut}} \frac{1}{2} \rho_\mathrm{ej} v_{\rm ej}^2 4 \pi r^2 \, dr\), where \(E_\mathrm{ej}\) is the energy of the ejecta.

At the injection time step \(t_\mathrm{SN}\), the grid cells within the insertion radius are updated. The total new density, \(\rho_\mathrm{new}\), is the sum of the ambient density, \(\rho_\mathrm{amb}\), and the ejecta density
\begin{equation}
    \rho_\mathrm{new} = \rho_\mathrm{amb} + \rho_\mathrm{ej}\,.
\end{equation}
To conserve momentum during the injection into the ambient medium, the new velocity, \(\mathbf{v}_\mathrm{new}\), of the gas in each cell is calculated as the mass-weighted average
\begin{equation}
    \mathbf{v}_\mathrm{new} = \mathbf{v}_\mathrm{amb} \frac{\rho_\mathrm{amb}}{\rho_\mathrm{new}} + \mathbf{v}_\mathrm{ej} \frac{\rho_\mathrm{ej}}{\rho_\mathrm{new}}\,.
\end{equation}

\subsection{Thermal conduction and cooling}
\label{sec:conduction}

Thermal conduction is classically described by the equation
\begin{equation}
    \frac{\partial T}{\partial t} = -\nabla \cdot \mathbf{q}_\mathrm{c} = \nabla \cdot \kappa \nabla T\,,
\end{equation}
where \(\kappa\) is the coefficient of conductivity. The coefficient for hydrogen plasma is given by \citet{spitzer_physics_1962}
\begin{equation}
    \kappa \approx 4.5 \times 10^{13} \left(\frac{T}{\SI{e8}{\kelvin}}\right)^{\frac{5}{2}} \left( \frac{\ln{\Lambda}}{40} \right)^{-1} \si{\erg\,\second^{-1} \centi\meter^{-1} \kelvin^{-1}} \approx KT^\frac{5}{2}\,,
    \label{eq:spitzer}
\end{equation}
where \(\ln \Lambda\) is the Coulomb logarithm, which is a~weak function of density and temperature; we assume \(\ln \Lambda \approx 30\). \(K\) is a~constant. We do not include thermal conductivity from neutral atomic collisions. 
Conduction is only significant at the steep temperature gradients near the contact discontinuity (\(T > \SI{e4}{\kelvin}\)), where the gas is strongly ionized and heat transport is entirely dominated by electron thermal conduction (\(\kappa_\mathrm{e} \propto T^{5/2}\)). In the colder, neutral regions of the shell, neutral thermal conductivity (\(\kappa_\mathrm{n} \propto T^{1/2}\), \citet{1953ApJ...117..431P}) is small and its contribution to the energy equation is negligible \citep{el-badry_evolution_2019}.

The classical model of thermal conduction assumes that the electron mean free path is much smaller than the temperature gradient scale length. This assumption may fail at extremely sharp interfaces, such as the contact discontinuity. Furthermore, on a discrete numerical grid, unresolved sharp gradients can drive artificially enormous heat fluxes that could unphysically drain a cell's energy in a single time step. To ensure both physical validity and numerical stability, the maximum amount of energy that can be transported by the free electrons is given by the saturated flux \citep{cowie_evaporation_1977, balbus_evaporation_1982}
\begin{equation}
    q_\mathrm{max} \approx \frac{3}{2} \rho c_\mathrm{s,iso}^3\,,
\end{equation}
where \(c_\mathrm{s,iso} = \sqrt{{k_\mathrm{B}T}/(\mu m_\mathrm{H})}\) is the isothermal sound speed. Because \(q_\mathrm{max}\) depends on the local density and temperature, it is evaluated dynamically for every cell during the simulation. For representative conditions in the shocked wind (\(T \approx \SI{e7}{\kelvin}\) and \(n \approx \SI{0.01}{\centi\meter^{-3}}\)) \(q_\mathrm{max} \approx \SI{8e-4}{erg\,cm^{-2}\,s^{-1}}\). Our diagnostics indicate that the gas rarely reaches the strongly saturated regime, but the flux limit prevents unresolved gradients from producing unphysically large heat fluxes. To transition smoothly between classical and saturated conduction, we use harmonic interpolation. \(K\) is then replaced by its effective value \(K_\mathrm{eff}\) defined as
\begin{equation}
    \frac{1}{K_\mathrm{eff} T^{\frac{5}{2}}}=\frac{1}{K T^{\frac{5}{2}}} + \lvert \nabla T \rvert \frac{2}{3 \rho c_\mathrm{s,iso}^3}\,.
    \label{eq:keff}
\end{equation}

We consider radiative cooling under the assumption that the gas is optically thin. This means that all of the radiation leaves the computational domain without being reabsorbed.
The cooling rate is given by
\begin{equation}
    Q = n_\mathrm{H}^2 \frac{n_\mathrm{e}}{n_\mathrm{H}} \Lambda_N(T,Z)\,,
    \label{eq:Q}
\end{equation}
where \(n_\mathrm{H}\) is the number density of hydrogen nuclei, \(Z\) is the metallicity of the gas, and \(\Lambda_N(T, Z)\) is the cooling function normalized to \(n_\mathrm{e} n_\mathrm{H} V = \SI{1}{\centi\meter^{-3}}\).
The simplified chemical state used by the equation of state (Section \ref{sec:hydro}) is not used to determine the cooling rate. Instead, for gas with \(T > \SI{15800}{\kelvin}\), we use the \citet{schure_new_2009} cooling tables, which provide the temperature-dependent fractional ionizations, \(x_\mathrm{e} = n_\mathrm{e}/n_\mathrm{H}\), and the individual contributions of elements to the cooling function, assuming solar metallicity. For subsolar metallicities, we proportionally scale the contributions according to the abundances of the elements.
For \(T < \SI{e4}{\kelvin}\), we utilize the low-temperature extension tables provided by \citet{schure_new_2009}, which are based on the results of \citet{dalgarno_heating_1972}. In this regime, we fix the fractional ionization to \(x_\mathrm{e} = 10^{-4}\). We chose this value because most of the gas in our simulations below \SI{e4}{\kelvin}, including the accreting gas, is dense, with temperatures \(T \le \SI{100}{\kelvin}\) and therefore overwhelmingly neutral. In the transition region \(\SI{e4}{\kelvin} < T < \SI{15800}{\kelvin}\), we interpolate between the two tables.

We omit the latent heat of ionization and dissociation from our calculations. This removes thermal inertia and makes the gas cool faster through the phase transitions.

To prevent overcooling at the contact discontinuity due to numerical mixing, we impose a limit on the cooling rate. We describe this in detail in Appendix \ref{app:limit}.

\subsection{Black hole accretion}

To evaluate the potential growth of the compact remnant, we estimate the mass accretion rate. Because the spatial resolution of our grid is insufficient to directly resolve the accretion flow onto a stellar-mass BH, we apply the Bondi accretion model \citep{bondi_spherically_1952} to establish an upper limit for the mass that could be transferred into an accretion disk and eventually accreted onto the BH. This gives us the maximum accretion rate
\begin{equation}
    \dot{M} \leq \frac{4\pi \lambda (GM)^2 \rho}{c_\mathrm{s}^3}\,,
    \label{eq:bondi}
\end{equation}
where \(M\) is the mass of the accreting object, \(\rho\) is the density of the gas, and \(c_\mathrm{s}\) is the sound speed.
The dimensionless constant \(\lambda\) depends on the equation of state of the gas. We adopt the isothermal value of \(\lambda = \frac{1}{4} e^{3/2} \approx 1.12\) for two reasons. Physically, the accreting material in our models originates from dense, radiatively cooled gas where the cooling timescale is much shorter than the dynamical accretion timescale, allowing the gas to efficiently radiate away compressional heating as it falls inward. Moreover, the isothermal assumption yields the theoretical maximum value for \(\lambda\), compared to \(\lambda = 0.25\) which holds for adiabatic gas. This aligns with our goal to calculate the upper limit for the accretion rate.

Mass is dynamically removed from the innermost cells at this calculated Bondi rate and added to the central point mass to update the gravitational potential. To prevent numerical instability from vacuum formation, the removal is bounded by a density floor (\(\rho_\mathrm{low} = \SI{e-27}{\gram\,\centi\meter^{-3}}\)). During the mass removal, the temperature is held constant; consequently, pressure decreases proportionally. The resulting density gradient, together with gravity, helps funnel gas towards the center. In this study, we do not include radiative or kinetic feedback (e.g., jets) from the newly formed BH, assuming all accreting material simply crosses the event horizon without further interaction with the surrounding gas.

\subsection{Simulation setup}
\label{sec:setup}

We ran our models for two different chemical compositions of stars: the high-metallicity mean composition of the Milky Way and the low-metallicity composition of I~Zwicky~18, which is thought to be a local equivalent of a primitive galaxy. Each simulation begins with a cloud of a~given mass, \(M_\mathrm{c}\), particle density, \(n_\mathrm{c}\), and pressure, \(p_\mathrm{c} / k_\mathrm{B} = \SI{e4}{\kelvin\,\centi\meter^{-3}}\). At this value, gas with \(n = \SI{100}{\centi\meter^{-3}}\) has a temperature of \SI{100}{\kelvin}. However, this pressure is negligible compared to the pressure inside a bubble, and its value does not significantly affect the results. In 3D, we introduced a random white noise density perturbation (with an amplitude of \SI{30}{\percent} of \(\rho\)) to the initial conditions to break up the perfect spherical symmetry.

The initial density of the cloud is \(n_\mathrm{c} \in [100, 1000]\,\si{\centi\meter^{-3}}\). This range was chosen based on physical constraints: if the density is too low, the bubble is too large and the density of the gas near the BH is insufficient for significant accretion. Conversely, if the density of the cloud is too high, the whole bubble collapses. This might initially suggest that the entire cloud mass becomes available for accretion onto the BH; however, the shell of the bubble is sufficiently massive to form stars. This process is not included in our model, so we cannot draw any conclusions about this case. We end the simulations either \SI{3}{\mega\year} after the SN explosion or \SI{6}{\mega\year} after the direct collapse into a BH. Because current stellar evolution models cannot definitively predict which massive stars explode and which collapse directly (as discussed in Section \ref{sec:introduction}), we systematically simulate both outcomes for every star in our sample.

For 1D, spherically symmetric simulations, we use a uniform grid with a resolution of 4096 cells, with reflecting boundary conditions at \(r=0\) and diode boundary conditions at the outer edge, which allow gas to leave the domain but prevent inflow. For 3D simulations, we use a~Cartesian grid with diode boundary conditions at all edges.

The size of the computational domain was chosen to fit the entire bubble for the full duration of the simulation in all cases where re-accretion is physically significant. These models have zero mass and energy fluxes through the boundaries. While gas can exit the domain in runaway expansion scenarios, the re-accreted mass in these evacuated bubbles is essentially zero regardless of boundary interaction.

In 3D, the domain is decomposed into octree blocks, with each block consisting of \(8^3\) interior cells and 4 layers of guard cells. We perform the simulations at a uniform resolution of \(256^3\) cells. For direct collapse scenarios, we further refine the area around the BH after its formation to an equivalent resolution of \(512^3\). Finally, in the 3D models, we turn off cooling for the first \SI{50}{\kilo\year} to prevent numerical overcooling before the bubble structure fully forms.

We adopt a fiducial explosion model with \(M_\mathrm{ej} = \SI{5.58}{M_\odot}\) and \(E_\mathrm{ej} = \SI{9.12e50}{\erg}\). These values correspond to the explosion parameters of a \SI{60}{M_\odot} progenitor \citep{martinez-gonzalez_supernovae_2019}. While explosion properties naturally vary with stellar mass, we fix these parameters to facilitate a direct comparison across parameter space.

Although applying a fixed ejecta mass to stars ranging up to \SI{257}{M_\odot} might initially appear to be a simplification, the hydrodynamical outcomes in this specific scenario are binary and highly insensitive to the exact explosion parameters. This is driven by two key physical factors:
\begin{enumerate}
    \item In all explored environments, even low-energy ejecta easily possess enough momentum to cross the entire pre-existing bubble interior and evacuate all dense structures into the outer shell. Conversely, significantly heavier and more energetic ejecta are still unable to overrun the massive, cold outer shell.
    \item The mass of the hot gas already present inside the bubble prior to the explosion is typically at least an order of magnitude higher than the mass of the supernova ejecta. Consequently, the exact value of \(M_\mathrm{ej}\) contributes negligibly to the overall mass budget potentially available for re-accretion.
\end{enumerate}

To validate these physical arguments, we conducted an exploratory parameter study varying both the environment and the explosion parameters. We tested an \SI{80}{M_\odot} high-metallicity progenitor setup with ambient densities and gravitational energies (\(n_\mathrm{c}\), \(E_\mathrm{grav}\)) of (\SI{100}{\centi\meter^{-3}}, \SI{e50}{erg}), (\SI{200}{\centi\meter^{-3}}, \SI{2e50}{erg}), and (\SI{500}{\centi\meter^{-3}}, \SI{3e50}{erg}), as well as a \SI{100}{M_\odot} low-metallicity progenitor with (\SI{200}{\centi\meter^{-3}}, \SI{5e50}{erg}) and (\SI{700}{\centi\meter^{-3}}, \SI{e50}{erg}). Across these configurations, we tested five sets of SN parameters (\(M_\mathrm{ej}\), \(E_\mathrm{ej}\)): (\SI{5.58}{M_\odot},\SI{5e50}{erg}), (\SI{10}{M_\odot}, \SI{9.12e50}{\erg}), (\SI{15}{M_\odot}, \SI{9.12e50}{\erg}), (\SI{10}{M_\odot}, \SI{e52}{\erg}), and (\SI{15}{M_\odot}, \SI{e52}{\erg}).

In all cases, we found that the parameters of the SN do not qualitatively affect the re-accretion outcome. The only quantitative difference we found was that SN ejecta move faster through the bubbles' interiors if they have higher mass or energy. This can lead to slightly larger and more energetic hot gas regions because the dense structures that contribute to cooling are removed from the interiors of the bubbles faster.

\section{Results}
\label{sec:results}

\begin{table}[t!]
    \begin{center}
    \caption{Simulation parameters and their evaluated ranges.}
    \label{tab:props}
    \begin{tabular}{clrlc}
        \hline
        ~ & description & range & unit & \(N_\mathrm{val}\) \\
        \hline
        \(n_\mathrm{c}\) & cloud particle density & \SIrange[range-phrase = \text{--}]{e2}{e3}{} & \si{\centi\meter^{-3}} & \SIrange[range-phrase = \text{--}]{3}{8}{} \\
        \(E_\mathrm{grav}\) & cloud grav. energy & \SIrange[range-phrase = \text{--}]{e49}{8e51}{} & \si{erg} & \SIrange[range-phrase = \text{--}]{3}{13}{} \\
        \(M_*\) & stars' ZAMS mass & \SIrange[range-phrase = \text{--}]{40}{257}{} & \si{M_\odot} & \(5\) \\
        \(Z\) & metallicity & \SIrange[range-phrase = \text{--}]{0.016}{0.66}{} & \si{Z_\odot} & 2\\
        \hline
    \end{tabular}
    \end{center}

    \vspace{2mm}
    \textbf{Notes.} \(N_\mathrm{val}\) represents the number of values varied for each parameter while keeping the others fixed.
\end{table}

\subsection{1D simulations of stellar wind bubbles}

Table \ref{tab:props} shows the range of parameters used in our study. We chose to parameterize the cloud by its density, which is directly related to the free-fall time, and by its gravitational binding energy, \(E_\mathrm{grav}\), which can be directly compared to the energy injected by feedback processes. The initial radii of the clouds are roughly in the range of  \SIrange[range-phrase = \text{--}]{9}{100}{pc}.

\begin{figure*}[th]
    \includegraphics[width=\textwidth]{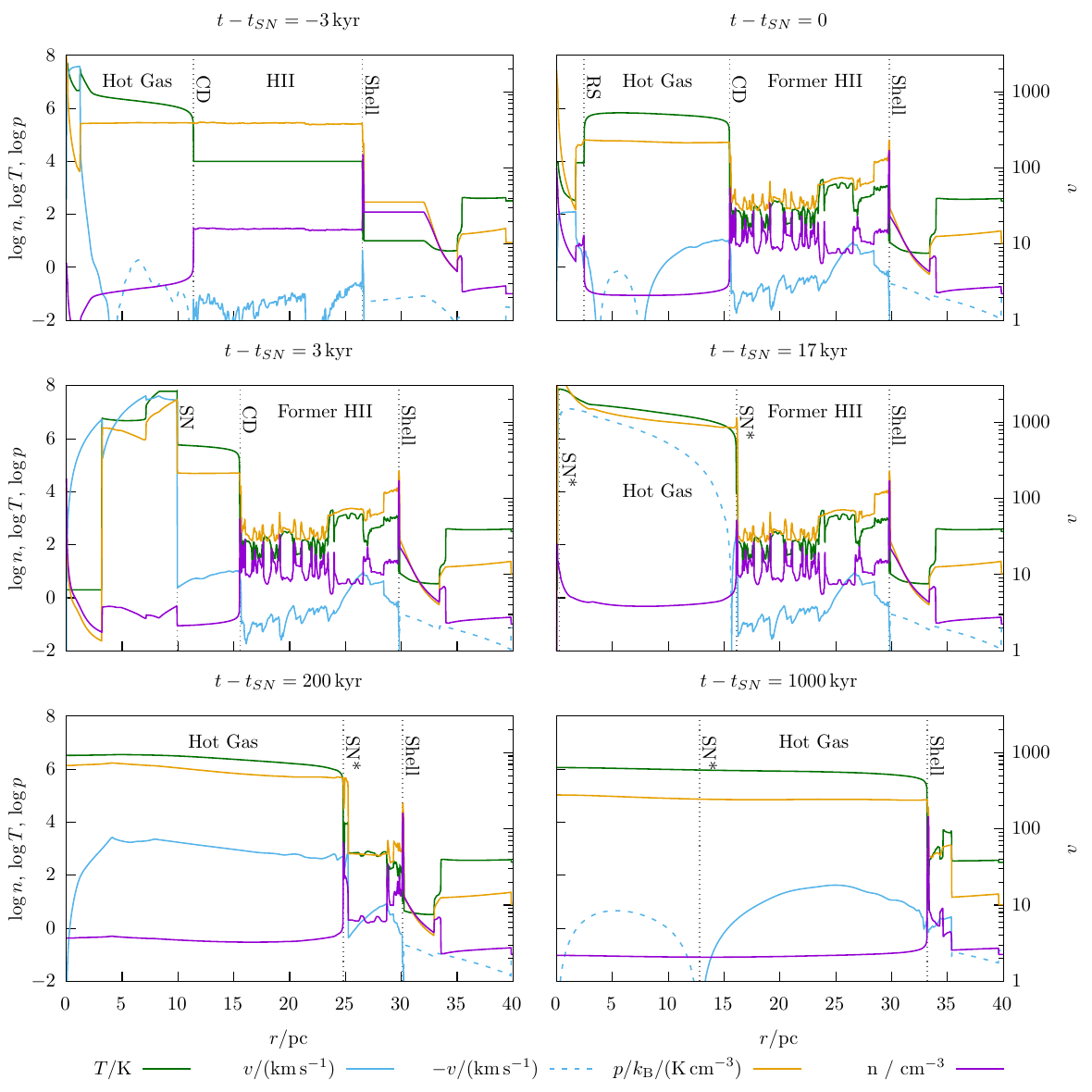}
    \caption{Profiles of temperature, \(T\), velocity, \(v\), pressure, \(p\), and particle density, \(n\) after an SN explodes inside an interstellar bubble. Negative velocity is shown using dashed lines. The bubble formed due to the wind from a \SI{77}{M_\odot} star with the metallicity of I Zwicky 18 inside a cloud with \(n_\mathrm{c} = \SI{100}{\centi\meter^{-3}}\) and \(E_\mathrm{grav} = \SI{e50}{erg}\). The panels show different times in the evolution of the system before and after the SN explosion at \(t_\mathrm{SN}\). Vertical dashed lines mark key radii: CD (contact discontinuity), RS (wind reverse shock), and SN (SN shock front). SN* indicate the reflected and transmitted shock fronts.}
    \label{fig:sn}
\end{figure*}

\begin{figure*}[t!]
    \centering
    \begin{minipage}{0.49\textwidth}
        \centering
        \(M_* = \SI{80}{M_\odot}, Z_* = \SI{0.66}{Z_\odot}\)
        \includegraphics[width=\textwidth]{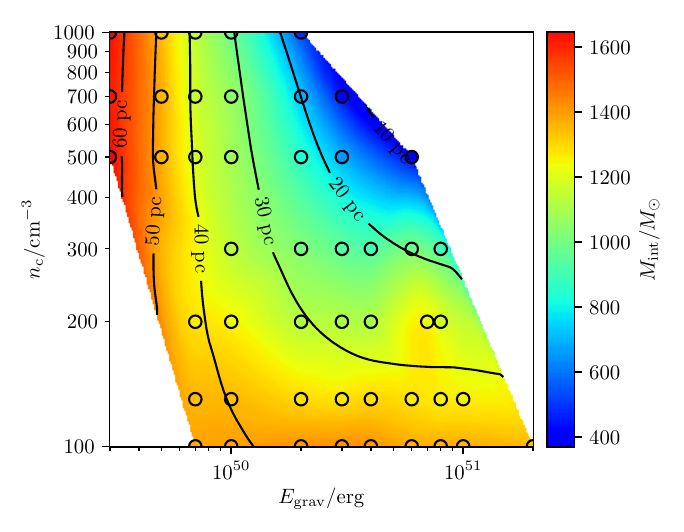}
        \begin{tikzpicture}[remember picture,overlay]
            \scriptsize
            \node[align=center] at (1.6,6.2) {\normalsize Bubble \\ \normalsize collapses \\ \normalsize before SN \\ \normalsize explosion};
            \node[align=center] at (-2.3,2.6) {\normalsize Runaway \\ \normalsize expansion};
        \end{tikzpicture}
    \end{minipage}%
    \hfill%
    \begin{minipage}{0.49\textwidth}
        \centering
        \(M_* = \SI{100}{M_\odot}, Z_* = \SI{0.016}{Z_\odot}\)
        \includegraphics[width=\textwidth]{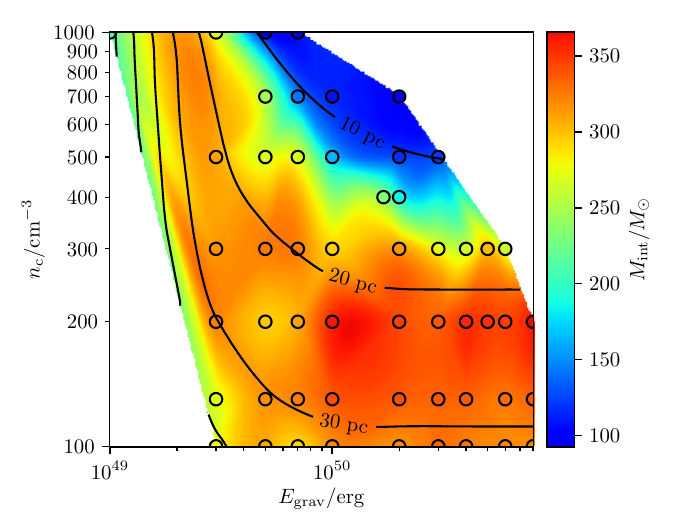}
        \begin{tikzpicture}[remember picture,overlay]
            \scriptsize
            \node[align=center] at (1.7,6.2) {\normalsize Bubble \\ \normalsize collapses \\ \normalsize before SN \\ \normalsize explosion};
            \node[align=center] at (-2.2,2.6) {\normalsize Runaway \\ \normalsize expansion};
        \end{tikzpicture}
    \end{minipage}
    \begin{minipage}{0.49\textwidth}
        \centering
        \vspace{-0.7cm}
        \includegraphics[width=\textwidth]{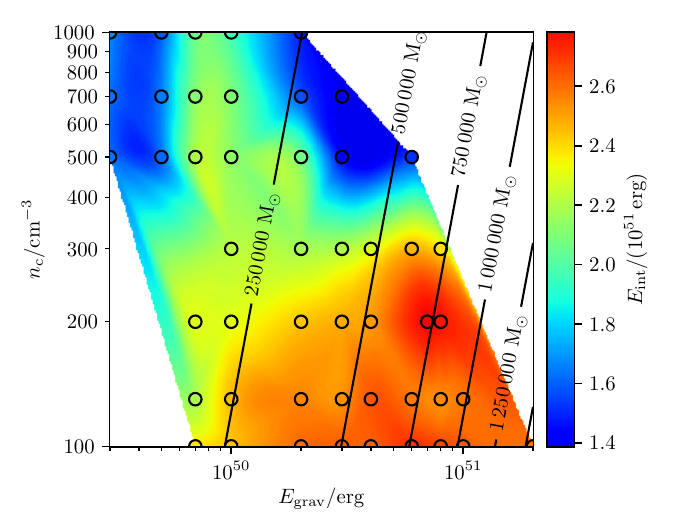}
    \end{minipage}%
    \hfill%
    \begin{minipage}{0.49\textwidth}
        \centering
        \vspace{-0.7cm}
        \includegraphics[width=\textwidth]{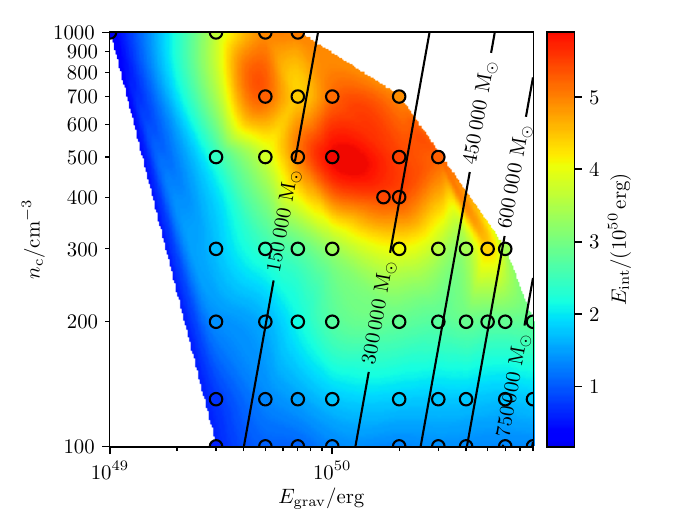}
    \end{minipage}
    \caption{Properties of the bubbles when the SN ejecta reflects from the dense shell as a function of the initial gravitational energy of the cloud, \(E_\mathrm{grav}\), and the initial particle density of the cloud, \(n_\mathrm{c}\). The top panels show the mass of the gas under the dense shell, \(M_\mathrm{int}\). The bottom panels show the internal energy of the gas inside the bubble excluding the shell, \(E_\mathrm{int}\). The contours show the radius of the shell and the initial mass of the cloud. The left column shows results for a \SI{80}{M_\odot} Milky Way metallicity star, the right column shows results for a \SI{100}{M_\odot} star with the metallicity of I~Zwicky~18. Black circles show individual simulations. Color maps and contours were calculated using cubic interpolation from the results of individual simulations. The white area in the top right corner of each plot corresponds to the high-density, high-gravity regime where the entire bubble and its shell gravitationally collapse back toward the star before the explosion (a scenario excluded from our analysis, as discussed in Section \ref{sec:setup}).}
    \label{fig:sn_maps}
\end{figure*}

\begin{figure*}[t!]
    \centering
    \begin{minipage}{0.49\textwidth}
        \centering
        \(M_* = \SI{77}{M_\odot}, Z_* = \SI{0.016}{Z_\odot}\)
        \includegraphics[width=\textwidth]{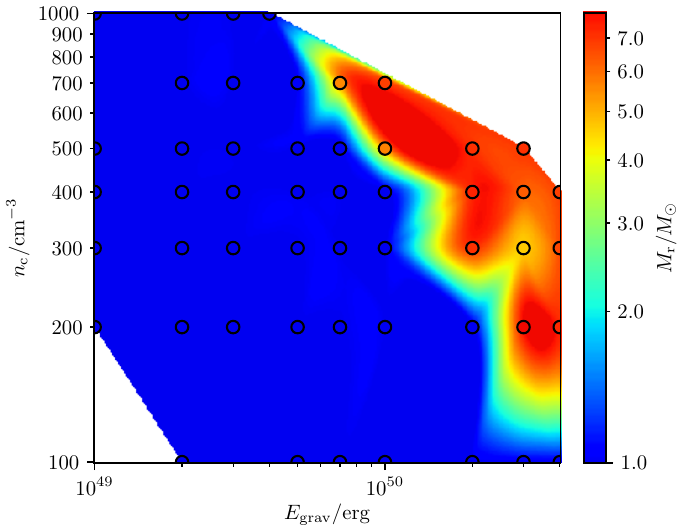}
        \begin{tikzpicture}[remember picture,overlay]
            \scriptsize
            \node[align=center] at (1.9,6.6) {\normalsize Bubble\\ \normalsize collapses\\ \normalsize before star};
        \end{tikzpicture}
    \end{minipage}%
    \hfill%
    \begin{minipage}{0.49\textwidth}
        \centering
        \(M_* = \SI{100}{M_\odot}, Z_* = \SI{0.016}{Z_\odot}\)
        \includegraphics[width=\textwidth]{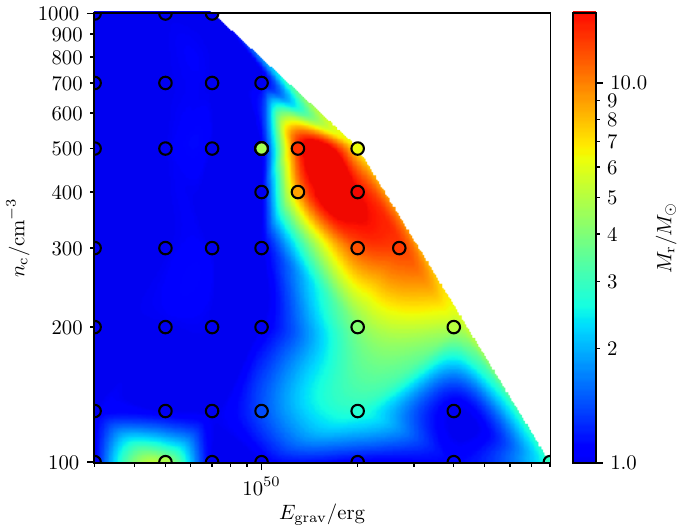}
        \begin{tikzpicture}[remember picture,overlay]
            \scriptsize
            \node[align=center] at (1.6,6.) {\normalsize Bubble\\ \normalsize collapses\\ \normalsize before star};
        \end{tikzpicture}
    \end{minipage}
    \caption{Total mass re-accreted onto a BH, \(M_\mathrm{r}\), formed by the direct collapse of a star as a function of the initial gravitational energy of the cloud, \(E_\mathrm{grav}\), and the initial particle density of the cloud, \(n_\mathrm{c}\). The plotted re-accreted mass represents the cumulative mass gained by the BH evaluated at the end of the simulation (\SI{3}{\mega\year} after the BH forms or immediately before the bubble collapse). The left column shows results for a~\SI{77}{M_\odot} progenitor, the right column shows results for a \SI{100}{M_\odot} progenitor, both with the metallicity of I Zwicky 18. Black circles show individual simulations. Color maps and contours were calculated using cubic interpolation in logarithmic space from the individual simulations.}
    \label{fig:nosn_reacc}
\end{figure*}

\begin{figure*}[t!]
    \centering
    \begin{minipage}{0.49\textwidth}
        \centering
        \(M_* = \SI{80}{M_\odot}, Z_* = \SI{0.66}{Z_\odot}\)
        \includegraphics[width=\textwidth]{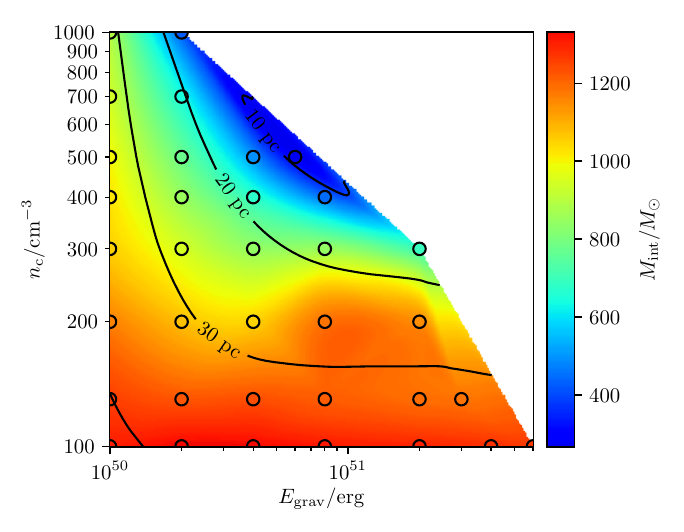}
        \begin{tikzpicture}[remember picture,overlay]
            \scriptsize
            \node[align=center] at (1.3,5.8) {\normalsize Bubble\\ \normalsize collapses\\ \normalsize before star};
        \end{tikzpicture}
    \end{minipage}%
    \hfill%
    \begin{minipage}{0.49\textwidth}
        \centering
        \(M_* = \SI{100}{M_\odot}, Z_* = \SI{0.016}{Z_\odot}\)
        \includegraphics[width=\textwidth]{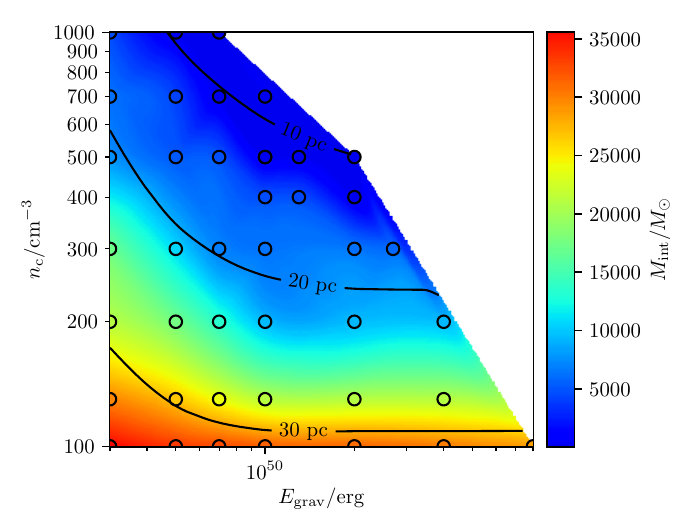}
        \begin{tikzpicture}[remember picture,overlay]
            \scriptsize
            \node[align=center] at (1.5,5.8) {\normalsize Bubble\\ \normalsize collapses\\ \normalsize before star};
        \end{tikzpicture}
    \end{minipage}
    \begin{minipage}{0.49\textwidth}
        \centering
        \vspace{-0.7cm}
        \includegraphics[width=\textwidth]{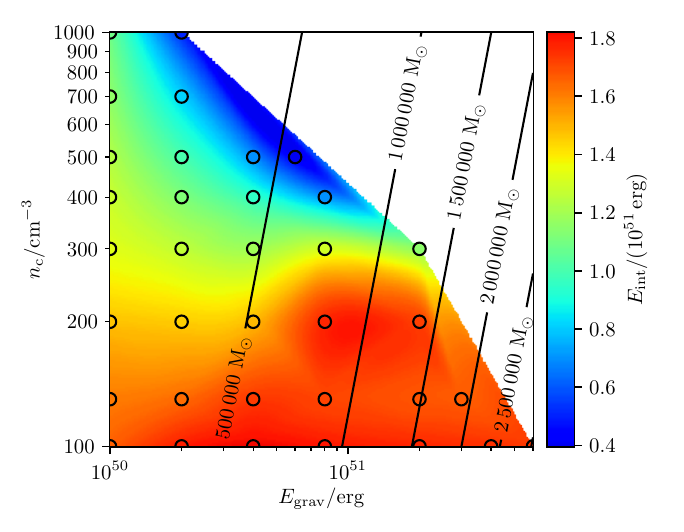}
    \end{minipage}%
    \hfill%
    \begin{minipage}{0.49\textwidth}
        \centering
        \vspace{-0.7cm}
        \includegraphics[width=\textwidth]{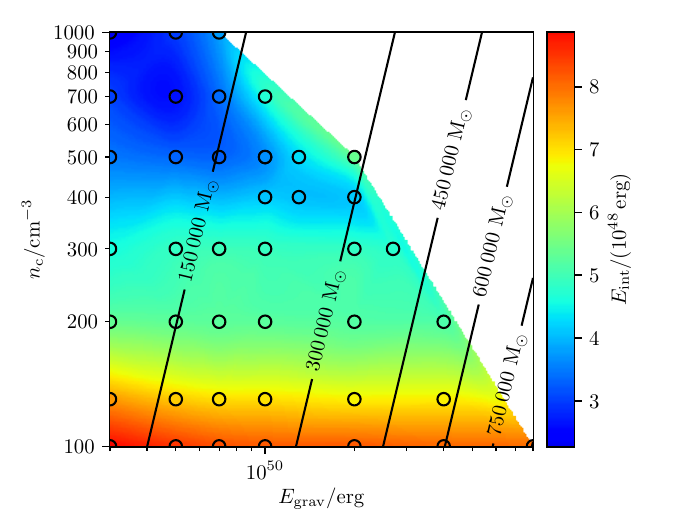}
    \end{minipage}
    \caption{Properties of the bubbles \SI{100}{\kilo\year} after the star collapses into a~BH as a function of the initial gravitational energy of the cloud, \(E_\mathrm{grav}\), and the initial particle density of the cloud, \(n_\mathrm{c}\). The top panels show the mass of the gas under the dense shell, \(M_\mathrm{int}\). The bottom panels show the internal energy of the gas inside the bubble excluding the shell, \(E_\mathrm{int}\), at the same time as the top panels. Contours show the radius of the shell and the initial mass of the cloud. The left column shows results for a~\SI{80}{M_\odot} Milky Way metallicity star, the right column shows results for a \SI{100}{M_\odot} star with the metallicity of I Zwicky 18. Black circles show individual simulations. Color maps and contours were calculated using cubic interpolation from the individual simulations.}
    \label{fig:nosn_maps}
\end{figure*}

\subsubsection{Black hole formation with a supernova}
\label{sec:1dsn}

We calculated 280 models differing in \(n_\mathrm{c}\), \(E_\mathrm{grav}\), \(Z\)~(2~values: Milky Way and I Zwicky 18), and \(M_*\) (5 values per metallicity), as summarized in Tables \ref{tab:stars} and \ref{tab:props}. To illustrate the evolution of an SN remnant, consider the case of a bubble formed by the wind of a \SI{77}{M_\odot} star with the metallicity of I Zwicky 18. This bubble formed within a~cloud with \(n_\mathrm{c} = \SI{100}{\centi\meter^{-3}}\) and \(E_\mathrm{grav} = \SI{e50}{erg}\).

Figure \ref{fig:sn} shows snapshots of the evolution before and after the SN explosion at \(t_\mathrm{SN}\). The first panel shows the bubble \SI{3}{\kilo\year} before the explosion. At the center is a fast stellar wind, surrounded by a hot shocked wind region. A contact discontinuity (marked CD) separates the hot gas from an \ion{H}{ii} region. A thin, dense shell (marked Shell) surrounds the bubble. It is important to note that the mass contained within the hot shocked wind region originates predominantly from the evaporation of the surrounding colder, denser gas (the \ion{H}{ii} region and the outer shell) driven by thermal conduction across the contact discontinuity. This evaporation dictates the mass budget of the interior. A detailed comparative analysis demonstrating these physical effects and showing the differences between simulations run with and without thermal conduction and radiative cooling is provided in Appendix \ref{app:limit} (see specifically Figure \ref{fig:convergence_bubble}).

The second panel shows the bubble just before the SN explosion. At the center, we see denser gas from the star's late evolutionary stages. The \ion{H}{ii} region has cooled and consists of a~medium-density, cold gas.
The third panel shows the SN ejecta passing through the shocked wind region (SN marks the shock front). The ejecta interact with the denser shells of the slow wind without significantly decelerating. The fourth panel shows the reflected wave reaching the BH at the center. However, the temperature is too high, so the pressure prevents most of the gas from accreting onto the BH. Instead, the wave is reflected and travels out. Meanwhile, as shown in the fifth panel, a transmitted wave passes through the former \ion{H}{ii} region and collects its mass. We mark both waves as SN*. The last panel shows the bubble after the transmitted wave has reached the dense outer shell. The transmitted wave is not powerful enough to pass through the shell, so it dissipates while a~weak reflected wave travels back toward the center.

Although a significant amount of gas approaches the BH\footnote{In the case shown in Figure \ref{fig:sn}, more than \(\SI{1}{M_\odot}\) of gas reaches the innermost \SI{2}{pc} during the first reflection.}, the pressure keeps it from being accreted, which holds true across our entire parameter space. However, an SN explosion significantly alters the bubble's structure. In all the cases we studied, the explosion was powerful enough to remove any dense structures from inside the bubble, yet not powerful enough to overcome the much heavier outer shell. This leaves us with bubbles filled with rare, hot gas that cannot cool efficiently.

Figure \ref{fig:sn_maps} shows the properties of the bubbles at the time when the SN ejecta reaches the cold, dense shell. At this point, the wave has swept up all dense structures within the bubbles and moved them into the shell.

For high-metallicity progenitors (e.g., the \(M_* = \SI{80}{M_\odot}\) Milky Way model shown in Figure \ref{fig:sn_maps}), we observe a strong correlation between \(M_\mathrm{int}\) and the overall size of the bubble. Because a~smaller bubble has a higher internal density for a~given enclosed mass, this compact configuration cools much more efficiently, suppressing the evaporation of material from the dense outer shell into the bubble's interior. A corresponding trend is evident for the internal energy, where \(E_\mathrm{int}\) approximately scales with bubble size across most of the parameter space. A notable deviation occurs in the regime of the largest bubble radii, where \(E_\mathrm{int}\) drops significantly. This is a direct consequence of the timescales involved in the shock propagation. The SN shock takes longer to traverse the cavity and reach the dense outer shell. This extended crossing time provides the hot, shocked gas with more time to undergo radiative cooling.

For low-metallicity progenitors (e.g., the \(M_* = \SI{100}{M_\odot}\) I~Zwicky~18 model shown in Figure \ref{fig:sn_maps}), the dependence of \(M_\mathrm{int}\) on the overall bubble size remains largely consistent with the high-metallicity case. However, the behavior of the internal energy, \(E_\mathrm{int}\), is flipped. This is a~result of differences in the circumstellar environment driven by metallicity. Unlike high-metallicity stars, low-metallicity stars emit more ionizing radiation and insert less energy through stellar winds during their main sequence phase. Furthermore, they lack the powerful Wolf--Rayet winds that, in high-metallicity stars, pressurize the interior, physically compressing the \ion{H}{ii} region into a thin layer that is completely swept into the cold outer shell once the star collapses. This results in a much broader \ion{H}{ii} region compared to the bubble size. The time it takes for the SN ejecta to cross the former \ion{H}{ii} region and reach the shell depends on its size. In some cases we explored, it can take up to approximately \(\SI{1}{\mega\year}\). The larger the bubble, the longer it has to cool before the wave reaches the shell.

Models utilizing different progenitor masses exhibit qualitatively similar behavior. The parameter space maps can be found in Appendix \ref{app:pspace}.

\subsubsection{BH formation with a direct collapse}
\label{sec:dc1d}

In this scenario, we analyze the evolution of a star that collapses directly into a BH without triggering an SN explosion. We calculated 222 models differing in \(n_\mathrm{c}\), \(E_\mathrm{grav}\), \(Z\) (2 values: Milky Way and I Zwicky 18), and \(M_*\) (5 values per metallicity). We identified four distinct cases with different outcomes; depending on the star's mass and metallicity, one of the following happens:

\begin{itemize}
\item High-mass, high-metallicity progenitors: Milky Way metallicity stars with masses \(M_* \gtrsim \SI{80}{M_\odot}\) experience a Wolf--Rayet phase with high mass-loss rates exceeding \(\SI{e-4}{M_\odot\per\year}\) with terminal wind velocities above \SI{5000}{\kilo\meter\per\second}. Such an amount of energy injected into the shocked wind region causes it to expand, compressing the \ion{H}{ii} region into a thin layer at the inner edge of the dense, cold shell. Similarly to an SN explosion, the fast wind clears the bubble of dense structures. High temperatures and pressures prevent a significant amount of gas accretion by the BH, and the gas does not cool due to its low density.

\item Low-mass, high-metallicity progenitors: Milky Way metallicity stars with masses \(M_* \lesssim \SI{60}{M_\odot}\) do not reach the Wolf--Rayet phase. Although they do experience strong late-stage mass loss, these winds lack the extreme velocities required to keep the bubble evacuated. Consequently, the dense \ion{H}{ii} region cools within approximately \(\SI{1}{\mega\year}\) and falls toward the BH. When the dense gas reaches the BH, tens of solar masses, and in some cases even hundreds of solar masses, are accreted onto the BH (see parameter space maps in Appendix \ref{app:pspace}). However, this requires perfect spherical symmetry. As we demonstrate in Section \ref{sec:3Ddirect}, more realistic 3D simulations show that the infalling gas does not reach the BH simultaneously, preventing the necessary cooling and shutting down accretion.

\item High-mass, low-metallicity progenitors: Lowering the metallicity of a star significantly modifies its evolution. The mass-loss rate through radiation-driven winds is much lower during the main sequence phase. Although these stars have much higher slow-wind mass-loss rates, they typically end their lives with much more mass. Additionally, during their main-sequence phase, lower-metallicity stars maintain much higher surface temperatures and therefore emit significantly more ionizing radiation. A star with the metallicity of I~Zwicky~18 and a mass \(M_* \gtrsim \SI{80}{M_\odot}\) ends its life as a red supergiant with high mass-loss rate reaching over \(\SI{e-4}{M_\odot\per\year}\), but with low wind velocity. This creates a massive shell that slowly moves outward through the bubble. The thermal pressure inside the bubble surrounding the inner shell confines it to the center and in some cases pushes the gas from the inner shell close to the BH. The maximum re-accreted mass depends on the progenitor mass, reaching \SI{7.85}{M_\odot}, \SI{15.3}{M_\odot}, \SI{17.1}{M_\odot}, and \SI{55.2}{M_\odot} for the \SI{77}{M_\odot}, \SI{100}{M_\odot}, \SI{150}{M_\odot}, and \SI{257}{M_\odot} models, respectively. As illustrated in the parameter space maps in Figure \ref{fig:nosn_reacc}, there is a~distinct region that allows for high re-accretion. For this to occur, the bubble must be in a state of collapse; however, the collapse must be gradual enough that the supergiant wind material is not swept up by the outer shell before it can reach the BH. The corresponding maps for the \SI{150}{M_\odot} (provided in Appendix \ref{app:pspace}) and \SI{257}{M_\odot} progenitors show a similar region of high \(M_\mathrm{r}\); however, we were unable to reproduce this in 3D (see Section \ref{sec:3Ddirect}). In one isolated simulation at low initial density and low gravitational energy (shown in Figure \ref{fig:nosn_reacc}), there is roughly \SI{5}{M_\odot} of re-accretion, where the pressure inside the bubble pushed some of the late stages of the wind towards the center. This does not happen in most simulations because of a~small amount of low-density warm gas close to the BH.

\item Low-mass, low-metallicity progenitors: A \SI{45}{M_\odot} star with low metallicity does not experience a~high mass-loss rate and does not create massive inner shells. The interior of the bubble remains warm (\SI{e4}{\kelvin}) and does not cool efficiently. The pressure of the interior prevents the dense gas from falling toward the BH and accreting.

\end{itemize}

For high-metallicity progenitors (e.g., the \SI{80}{M_\odot} Milky Way model shown in Figure \ref{fig:nosn_maps}), the post-collapse evolution is similar to the SN explosion scenario discussed in Section \ref{sec:1dsn}. The pressure of the hot gas forces the interior to expand into the former \ion{H}{ii} region, sweeping the dense gas into the shell. Only in the largest bubbles (those forming in clouds with low initial density and low gravitational binding energy) does the former \ion{H}{ii} region resist being entirely swept up, resulting in a much higher retained mass inside the bubble.

In contrast, low-metallicity progenitors (e.g., the \SI{100}{M_\odot} I~Zwicky~18 model in Figure \ref{fig:nosn_maps}) retain much of the mass inside the former \ion{H}{ii} region after the star collapses. Because this interior mass is significantly higher, the gas cools much more efficiently. Coupled with the weaker mechanical feedback, this efficient cooling results in a substantially lower total internal energy within the bubble.

The plots for other stars with different masses show similar trends. The parameter space maps can be found in Appendix \ref{app:pspace}.

\subsection{3D simulations of stellar wind bubbles}

\begin{figure*}
    \centering
    \includegraphics[width=\textwidth]{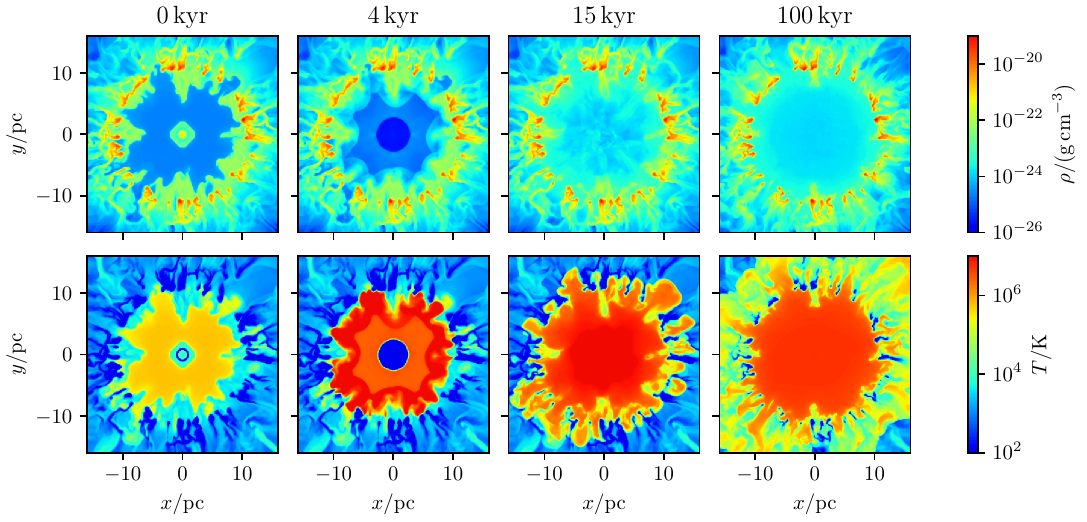}
    \caption{Density, \(\rho\), and temperature, \(T\), after an SN explosion inside an interstellar bubble. The bubble was formed by the wind from a \SI{77}{M_\odot} star with the metallicity of I Zwicky 18 inside a cloud with \(n_\mathrm{c} = \SI{500}{\centi\meter^{-3}}\) and \(E_\mathrm{grav} = \SI{5e49}{erg}\). The panels show cross-sectional slices in the \(z=0\) plane at different times. From left to right: immediately after the SN explosion; when the SN wave impacts the former \ion{H}{ii} region; after the reflected wave reaches the central BH; and finally, \SI{100}{kyr} after the explosion.}
    \label{fig:SN3D}
\end{figure*}

\begin{table}[t]
    \begin{center}
    \caption{Initial parameters and re-accreted masses in 3D.}
    \label{tab:3D}
    \begin{tabular}{lll|lll}
    \hline
        \(M_* / M_\odot\) & \(n_\mathrm{c} / \si{\centi\meter^{-3}}\) & \(E_\mathrm{grav} / \si{\erg}\) & \(M_\mathrm{r,1D} / \si{M_\odot}\) & \(M_\mathrm{r,3D} / \si{M_\odot}\) \\ \hline
        \(60\) & \(100\) & \SI{2e50}{} & \SI{0.6}{} & \SI{2e-6}{} \\
        \(60\) & \(200\) & \SI{5e50}{} & \SI{31}{} & \SI{2e-6}{} \\
        \(77\) & \(400\) & \SI{3e50}{} & \SI{6.2}{} & \SI{4.0}{} \\
        \(77\) & \(500\) & \SI{5e49}{} & \SI{0.2}{} & \SI{4e-4}{} \\
        \(77\) & \(500\) & \SI{2e50}{} & \SI{7.0}{} & \SI{3.2}{} \\
        \(100\) & \(300\) & \SI{2e50}{} & \SI{11.6}{} & \SI{6.9}{} \\
        \(100\) & \(500\) & \SI{1e50}{} & \SI{4.6}{} & \SI{2.1}{} \\
        \(150\) & \(300\) & \SI{2e50}{} & \SI{17.1}{} & \SI{2.8}{} \\
        \(257\) & \(200\) & \SI{4e50}{} & \SI{55.2}{} & \SI{0.1}{} \\ \hline
    \end{tabular}
    \end{center}

    \vspace{2mm}
    \textbf{Notes.} Columns represent the progenitor zero-age main-sequence mass, \(M_*\), initial cloud particle density, \(n_\mathrm{c}\), and initial gravitational energy of the cloud, \(E_\mathrm{grav}\). The final two columns show the re-accreted mass for the direct collapse case in 1D, \(M_\mathrm{r,1D}\), and 3D, \(M_\mathrm{r,3D}\), respectively.
\end{table}

To verify our 1D results, we selected nine representative 3D simulations (Table \ref{tab:3D}) exploring both the SN explosion and direct collapse scenarios. Due to computational costs, these run at an \(8\times\) lower spatial resolution than the 1D parameter study. Low-metallicity progenitors with initial masses \(\gtrsim \SI{80}{M_\odot}\) are of particular interest because they create bubbles with vast \ion{H}{ii} regions containing dense gas inside the bubble.

\subsubsection{Black hole formation with a supernova}

The re-accretion outcome of all of the simulations is consistent with the results of the high-resolution 1D simulations: no re-accretion occurs when an SN is included.

For the high-metallicity progenitor with an initial mass of \SI{60}{M_\odot}, the overall structure and evolution of the bubble are consistent with the 1D counterparts. No re-accretion occurred, even in the highly compact case with \(n_\mathrm{c} = \SI{200}{\centi\meter^{-3}}\), where the radius of the bubble at the time of the SN explosion is only \SI{6}{pc}.

For low-metallicity progenitors, there are two differences. Near the end of the star's life, when the \ion{H}{ii} region cools (if present), it does not maintain spherical symmetry; instead, it becomes unstable, collapsing into dense, filamentary structures that penetrate from the shell into the bubble's interior. Moreover, in some cases, the shell becomes Rayleigh--Taylor unstable and fractures.

The evolution of a low-metallicity SN remnant (e.g., the \SI{77}{M_\odot} model shown in Figure \ref{fig:SN3D}) illustrates the interaction between the SN ejecta and the structured circumstellar medium. Following the explosion, the SN shock propagates outward and impacts the former \ion{H}{ii} region, generating a reflected wave that travels back toward the central BH. The density near the BH remains too low to trigger efficient cooling. The resulting pressure prevents any significant accretion.

\subsubsection{Black hole formation with a direct collapse}
\label{sec:3Ddirect}

\begin{figure*}
    \centering
    \includegraphics[width=\textwidth]{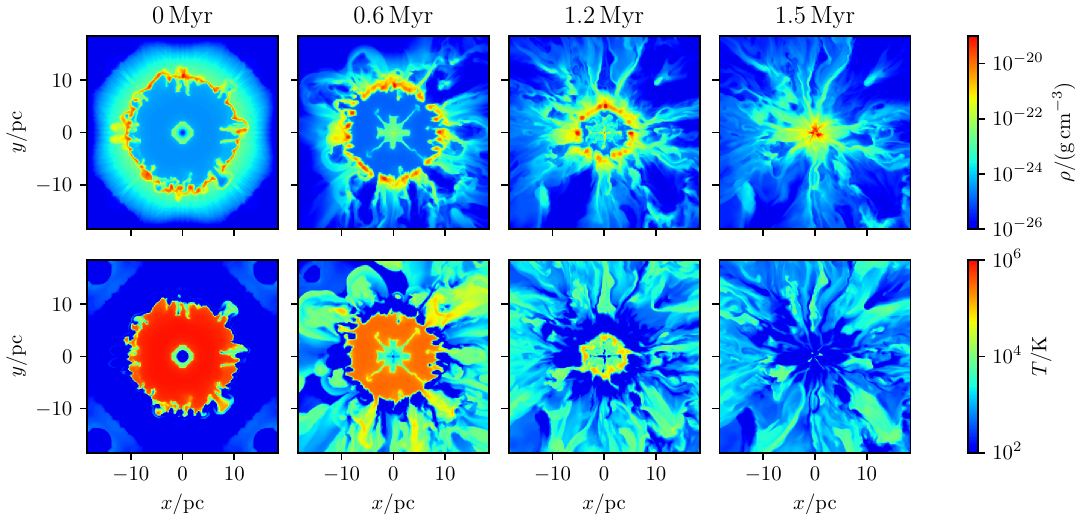}
    \caption{Density, \(\rho\), and temperature, \(T\), after a direct collapse inside an interstellar bubble. The bubble was formed by the wind from a \SI{100}{M_\odot} star with the metallicity of I Zwicky 18 inside a cloud with \(n_\mathrm{c} = \SI{500}{\centi\meter^{-3}}\) and \(E_\mathrm{grav} = \SI{e50}{erg}\). The panels show cross-sectional slices in the \(z=0\) plane at different times after the star collapses to a BH. From left to right: shortly after the star collapses into a BH; during the onset of the bubble's collapse; as the dense central wind material cools and accretes; and shortly before the complete collapse of the bubble.}
    \label{fig:noSN3D}
\end{figure*}

Table \ref{tab:3D} shows the re-accreted masses from our 1D and 3D simulations for all explored cases. There is a major difference for the high-metallicity \SI{60}{M_\odot} progenitor. While our 1D models predicted massive re-accretion, our 3D simulations produce negligible accretion. This difference arises from the relaxation of forced spherical symmetry. In 1D, spherical symmetry causes infalling gas to arrive at the center simultaneously, creating a density spike that rapidly cools to \SI{100}{\kelvin} and triggers accretion. In contrast, 3D simulations naturally break this symmetry; tiny deviations cause the gas to arrive at slightly different times, preventing densities required for fast cooling.

For low-metallicity progenitors, we confirm that re-accretion still happens in the 3D simulations. However, because the gas does not fall in perfectly symmetrically, the final BH mass is usually lower than what we measured in the 1D models.

The evolution following a direct collapse (e.g., the \SI{100}{M_\odot} model shown in Figure \ref{fig:noSN3D}) highlights the critical role of late-stage stellar winds in feeding the central BH. Shortly after the star collapses, dense material ejected during the progenitor's slow-wind phase remains in the inner region. As stellar feedback decreases, the bubble begins to contract. Initially, the hot interior maintains enough thermal pressure to confine this dense wind material to the center. As the contraction progresses, this dense central gas cools efficiently down to \SI{100}{\kelvin}. The reduced pressure makes accretion onto the BH possible. By the time the outer shell completely collapses, the BH has accreted up to \SI{6.9}{M_\odot} of gas, yielding a final BH mass of up to \SI{93}{M_\odot} for the \SI{100}{M_\odot} progenitor, and \SI{73}{M_\odot} for the \SI{77}{M_\odot} progenitor.

In the case of the \SI{150}{M_\odot} and \SI{257}{M_\odot} progenitors, we were unable to reproduce the high re-accretion observed in 1D simulations. In 1D, the dense material from the late, slow wind is confined to the center and does not get outside the inner few parsecs. In contrast, the relaxation of forced spherical symmetry in 3D allows the dense gas to mix with the surrounding hot medium, thereby reducing the central density below the threshold required for significant accretion. Furthermore, the core of the \SI{150}{M_\odot} star has roughly \SI{83}{M_\odot}, which falls into the pair-instability range.

These 3D results are broadly consistent with our 1D simulations, although the total accreted mass is smaller in 3D compared to 1D. This minor discrepancy, along with differences in the accretion dynamics, arises from several structural and numerical factors. Primarily, accretion in the 1D models occurs in discrete episodes driven by perfectly spherical waves reflecting off the center and compressing the gas. In 3D, natural deviations from strict spherical symmetry prevent the wavefronts from reaching the center all at once, resulting in a more gradual accretion process (see Appendix \ref{app:accretion_comparison} for a comparison). Furthermore, the dynamics of late-stage collapse in our idealized model may be heavily influenced by the bubble's shell itself, whose structure would likely be altered by local star formation, which is not included in this model. Finally, the comparatively lower resolution of the 3D simulation leads to slightly faster cooling, which can moderately overestimate the accretion rate.

\section{Discussion}
\label{sec:discussion}

\subsection{Comparison with previous studies}

\citet{tenorio-tagle_evolution_1990, tenorio-tagle_evolution_1991, dwarkadas_evolution_2005, martinez-gonzalez_supernovae_2019} performed similar simulations of SN explosions inside wind-blown bubbles. Our results are consistent with their findings. We observed the same interaction between the SN ejecta and the structures inside the bubble that create reflected and transmitted waves. However, we explored a~much wider range of parameters. We demonstrate that the shell of the bubble is massive enough to stop the wave caused by the SN explosion in all of the studied cases.

\subsection{Astrophysical implications}

Our results reveal that re-accretion is a physically viable channel for stellar-mass BH growth in low-metallicity direct-collapses, although the allowed parameter space is restricted. In particular, re-accretion of up to \SI{6.9}{M_\odot} can occur when a low-metallicity star collapses directly into a BH without an SN explosion. In contrast, re-accretion does not occur when the star explodes as an SN or in the case of a high-metallicity progenitor. Fast stellar winds from high-mass, high-metallicity stars, as well as SN explosions, effectively clear the bubble of dense structures, leaving behind a hot, high-pressure, and low-density interior that cannot be efficiently accreted.

As discussed in Section \ref{sec:1dsn} and demonstrated by the comparative models in Appendix \ref{app:limit}, thermal conduction plays a~significant role in determining the structure and evolution of interstellar bubbles. The evaporation of shell mass into the interior increases the mass of the hot gas and decreases both the energy and the size of a bubble. Another key factor is considering the entire evolution of the star. Fast stellar winds from high-mass, high-metallicity stars efficiently clear the bubble of dense structures. In contrast, the strong ionizing radiation of low-metallicity stars heats the cold gas at the inner boundary of the shell, causing it to expand inward and fill the bubble's interior with a vast, relatively dense \ion{H}{ii} region. Furthermore, an SN explosion acts similarly to fast stellar winds by pushing most of the gas from the bubble into the cold, dense shell.

Massive stars---particularly luminous blue variables like \(\eta\)~Carinae---can undergo giant eruptions that eject a~substantial amount of mass (\(\gtrsim \SI{10}{M_\odot}\)) in a short-duration event, creating a~dense, compact circumstellar shell \citep{smith_mass_2014}. If an SN explosion occurs shortly after such an eruption, the interaction between the ejecta and this dense shell can become strongly radiative, significantly enhancing the cooling efficiency \citep{serrano-hernandez_bright_2025}. 
While our current study indicates that re-accretion is inefficient in hot, low-density bubbles, the rapid cooling triggered by the collision with a~massive eruptive shell could theoretically lower the internal pressure enough to facilitate accretion. Future studies should explore whether concentrating the mass loss into discrete, dense shells alters the accretion budget in these extreme scenarios.

Furthermore, while this study focuses on the mass growth of stellar-mass BHs, similar accretion mechanisms hold broader astrophysical relevance for other isolated compact remnants. For instance, long-term fallback accretion of SN or circumstellar material onto newly formed isolated neutron stars has been proposed as a potential energy source to explain the persistent X-ray emission observed in Anomalous X-ray Pulsars \citep{2013ApJ...764...49T}. Although the compact object and observational signatures differ, the underlying hydrodynamic constraints governing the late-time fallback of material remain closely related.

Ultimately, re-accretion offers a promising new piece of the puzzle regarding the pair-instability mass gap. While a post-formation mass increase of up to \SI{6.9}{M_\odot} is insufficient to fully bridge the gap, it pushes the upper boundary of BH masses achievable through isolated single-star evolution.

\subsection{Model limitations and future prospects}
\label{sec:limitations}

To explore a wide parameter space, our 1D parameter study requires the assumption of spherical symmetry. The initial conditions of our 3D simulations assumed a spherical, globally homogeneous cloud with cell-scale white noise perturbations, but without pre-existing turbulence embedded in a low-density ambient medium. Consequently, we did not observe localized effects, such as the formation of dense clumps that could fall through the high-pressure gas toward the BH. Introducing inhomogeneities and driven turbulence into the ambient cloud could alter the cooling times and accretion rates, warranting further investigation.

Evolving symmetric initial conditions on a Cartesian mesh with a directionally split hydrodynamics solver seeds instabilities along the principal grid axes, a numerical artifact related to the carbuncle phenomenon \citep[e.g.,][]{1994IJNMF..18..555Q}. Furthermore, as the system evolves, the low-density ambient gas from the corners of the computational domain falls inward, creating an odd visual distribution outside the cloud. However, the mass involved in these low-density corners is entirely negligible compared to the mass of the dense shell (which contains almost the entire mass of the initial cloud), meaning its dynamical influence on the bubble interior is insignificant.

We use a simplified prescription for \(\mu\), which is calculated only as a function of temperature. To verify that our results are not sensitive to this, we recalculated the 1D parameter space study for the case with the \SI{100}{M_\odot} star. We performed two test cases: one without any molecular gas and one with only ionized gas. In both cases, the outcomes did not change significantly. The parameter space plots are shown in Appendix~\ref{app:mu}.

Additionally, our model omits star formation within the dense shell of the swept-up ISM. The shell's density is theoretically sufficient to trigger star formation. However, star formation in such shellular geometries may yield a highly bottom-heavy initial mass function \citep{2018A&A...612A..55S}. The resulting low-mass stars would generate negligible mechanical and radiative feedback. Furthermore, their condensation timescales would far exceed our few-megayear post-collapse window, meaning the omission of local star formation would not alter our accretion limits.

Another assumption we make is that the BH forms with zero velocity. This may not be true for an asymmetric SN explosion \citep{shklovskii_possible_1970, janka_natal_2013, wongwathanarat_three-dimensional_2013}, BH natal kicks \citep{bonnell_gravitational_1995}, or asymmetric neutrino emission during core collapse \citep{gourgoulhon_upper_1993, fryer_effects_2006}. Consequently, the BH would spend only a limited amount of time inside the bubble, making re-accretion more unlikely.
However, population studies of globular clusters indicate that BH natal kicks must generally be much lower than those of neutron stars to explain the observed cluster retention fractions \citep{2018A&A...617A..69P}, allowing dense BH subsystems to remain bound and evolve dynamically. Furthermore, this is significantly less problematic for the direct collapse scenario, which is the primary channel where we observe substantial re-accretion. Theoretical models predict that stars undergoing direct collapse without an explosion eject negligible baryonic mass, resulting in near-zero hydrodynamic natal kicks \citep{mandel_simple_2020}. The only recoil arises from asymmetric neutrino emission, which is expected to impart minimal velocities. This has been observationally supported by recent measurements of binary systems like VFTS 243, where the BH is inferred to have formed via complete collapse with a natal kick of \(\lesssim \SI{10}{\kilo\meter\per\second}\) \citep{vigna-gomez_constraints_2024}. Therefore, the assumption of a stationary BH remains physically robust for the direct collapse pathways that favor re-accretion.

We also limited our study to single stars that evolve in isolation. However, a large majority of massive stars are found in multiple stellar systems \citep{duchene_stellar_2013, sota_galactic_2014}. A binary companion could significantly alter the evolutionary tracks, the mass-loss rates, and the resulting structure of the wind-blown bubble.

\section{Conclusions}
\label{sec:conclusions}

We performed 1D and 3D hydrodynamic simulations of interstellar bubbles driven by feedback from massive stars. These simulations included saturated thermal conduction, radiative cooling, ionizing UV radiation, gravity of the star and BH, self-gravity of the gas, stellar winds, SN explosions, and gas accretion onto newly formed BHs.

We followed the evolution of young massive stars inside dense clouds. We investigated two scenarios: the star forming a stellar-mass BH in a core-collapse SN, and the star directly collapsing to a BH. We explored two metallicities: Milky Way and I Zwicky 18. We examined how much mass could be accreted from the pre-blown bubble onto the stellar-mass BH over millions of years after its formation. Within the parameter space explored here, our key findings are as follows:

\begin{enumerate}
  \item Low-metallicity stars with masses \(\gtrsim \SI{77}{M_\odot}\) end their lives with slow, high-mass-loss winds that form massive inner shells. Re-accretion of up to \SI{6.9}{M_\odot} of the inner shell gas is possible after a low-metallicity progenitor directly collapses into a BH, resulting in final BH masses of up to \SI{93}{M_\odot}.
  \item An SN explosion inside a bubble injects significant energy into the hot gas and moves all dense structures into the shell. Consequently, the interior of the bubble consists of gas that is too hot (\SI{e4}{\kelvin}, which implies sound speeds \(\gtrsim \SI{10}{\kilo\meter\per\second}\) and a small Bondi radius) to be efficiently accreted and not dense enough for efficient cooling.
  \item Fast, strong winds from massive high-metallicity stars with masses \(\gtrsim \SI{80}{M_\odot}\) act similarly to SN explosions, preventing significant accretion.
  \item Thermal conduction and stellar feedback mechanisms (radiative cooling and ionizing UV radiation) fundamentally govern the mass budget of the bubble interior. Specifically, conduction drives the evaporation of shell material, while ionizing radiation dictates the presence of dense \ion{H}{ii} regions, both of which must be included in realistic simulations.
  \item Re-accretion represents a promising new physical pathway for stellar-mass BH growth. By enabling post-formation mass gains of several solar masses, this mechanism helps to narrow the pair-instability mass gap and brings us closer to explaining the unexpectedly massive BHs detected in recent gravitational wave mergers.
\end{enumerate}

\begin{acknowledgements}

We thank the anonymous reviewer for their constructive comments and suggestions that helped improve the clarity and rigor of this manuscript.

D.K., R.W., and S.J. acknowledge the support by the Czech Ministry of Education, Youth and Sports, through the INTER-EXCELLENCE II program, project LUC24023 (MSMT-14950/2024-5), and by the project RVO:67985815.

D.K. acknowledges the support from the Grant Agency of Charles University project No. 170226.

D.S. acknowledges that this research was funded in part by the National Science Center (NCN), Poland under grant number OPUS 2021/41/B/ST9/00757. For the purpose of Open Access, the authors have applied a CC-BY public copyright license to any Author Accepted Manuscript (AAM) version arising from this submission.

The software used in this work was developed in part by the DOE NNSA- and DOE Office of Science-supported Flash Center for Computational Science at the University of Chicago and the University of Rochester.

\end{acknowledgements}

\bibliographystyle{aa}
\bibliography{bibliography}

%
%

\begin{appendix}

\section{Velocity of the wind}
\label{app:vinf}

During the main sequence phase, we calculate the terminal velocity of the wind, \(v_\infty\), using the relation by \citet{krticka_new_2021}
\begin{equation}
    v_\infty = \frac{1}{2} \left [ (v_+-v_-) \frac{T_\mathrm{eff}-T_0}{\Delta T(1+\lvert v \rvert)} + v_+ + v_- \right ] v_\mathrm{esc}\,,
    \label{eq:vinfty_krticka}
\end{equation}
where \(v_+ = 3.0\), \(v_- = 0.7\), \(T_0 = \SI{20700}{\kelvin}\), and \(\Delta T = \SI{3300}{\kelvin}\) are constants. The escape velocity, \(v_\mathrm{esc}^2 = {2 G M_*}/{R_*}\), is calculated from the tracks, \(G\) is the universal gravitational constant. Equation \ref{eq:vinfty_krticka} holds for solar metallicity, \(Z_\odot\). To account for different chemical compositions, we scale the terminal velocity using the metallicity dependence derived by \citet{leitherer_deposition_1992}. This yields the final expression for the terminal velocity
\begin{equation}
    v_\infty = \frac{1}{2} \left [ (v_+-v_-) \frac{T_\mathrm{eff}-T_0}{\Delta T(1+\lvert v \rvert)} + v_+ + v_- \right ] v_\mathrm{esc} \left (\frac{Z}{Z_\odot} \right )^{0.13}\,,
\end{equation}
where \(Z\) is the metallicity of the star.

For red supergiants, we set \(v_\infty = \SI{30}{\kilo\meter\per\second}\) \citep{airapetian_winds_2010}, and for luminous blue variable stars, \(v_\infty = \SI{200}{\kilo\meter\per\second}\) \citep{vink_fast_2018}. During the Wolf--Rayet phase, we use a relation from \citet{sander_nature_2020}
\begin{equation}
    \log \frac{\dot{m}}{v_\infty} = k \log [-\log (1-\Gamma_\mathrm{e})] + l\,,
    \label{eq:vinfty_sander}
\end{equation}
where \(k = 2.5\) and \(l=-6.3\) are constants, \(\Gamma_\mathrm{e}\) is the Eddington factor which is given in the evolution tracks, and the empirical relation assumes \(\dot{m}\) is in units of \si{M_\odot\,\year^{-1}} and \(v_\infty\) is in \si{\kilo\meter\,\second^{-1}}.

\begin{figure}[h]
    \centering
    \includegraphics[width=\linewidth]{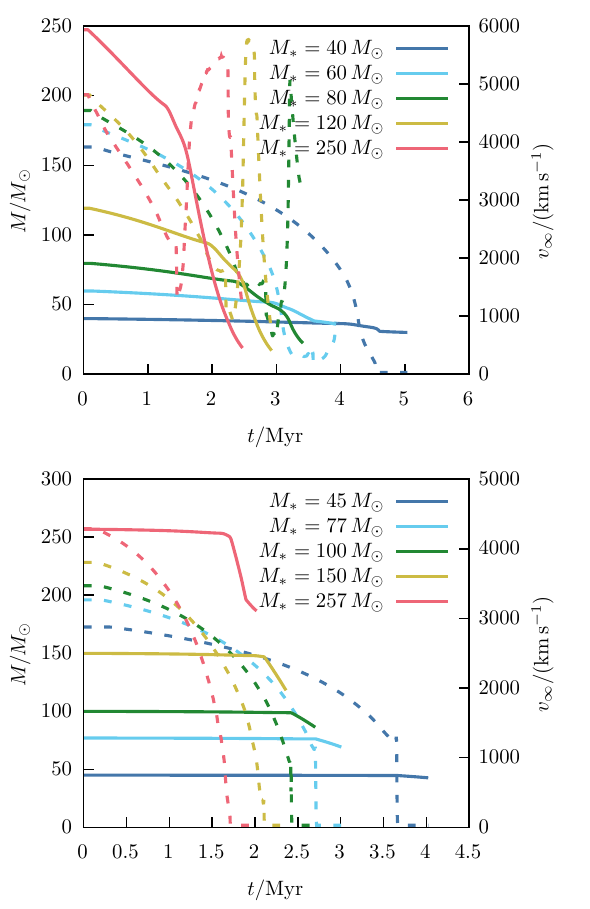}
    \caption{Mass of the stars, \(M\), and the terminal velocity of their winds, \(v_\infty\), as a function of time, as calculated from the Bonn Optimized Stellar Tracks \citep{szecsi_bonn_2022}. The top panel shows stars with \(Z = \SI{0.66}{Z_\odot}\) and the bottom panel shows stars with \(Z = \SI{0.016}{Z_\odot}\). \(M_*\) is the zero-age main sequence mass of a star.}
    \label{fig:tracks}
\end{figure}

Figure \ref{fig:tracks} shows the masses of stars and the terminal velocities of their winds as a function of time.

\section{Cooling limit}
\label{app:limit}

Contact discontinuities pose a significant problem for numerical solvers because they require extremely fine resolution to capture steep profiles accurately. Thermal conduction smooths out the temperature and density gradients, creating cells with temperatures between those of the two media. The cooling rate is a~strong function of temperature and density.

\begin{figure}[htb]
    \centering
    \includegraphics[width=\linewidth]{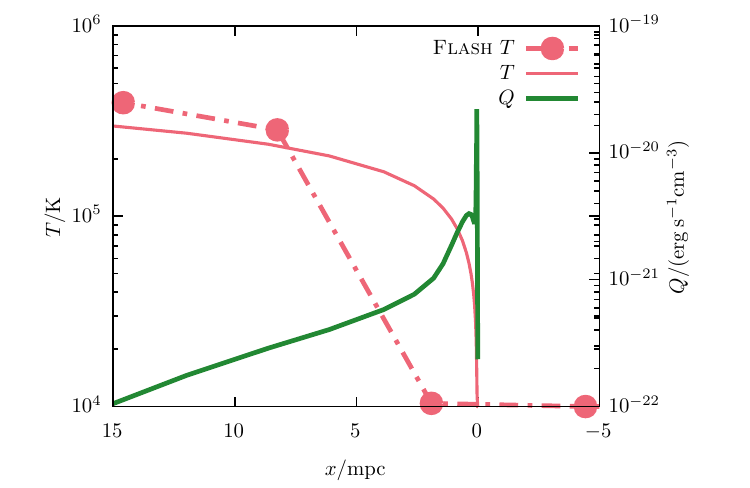}
    \caption{Zoomed-in temperature profile, \(T\), calculated using \textsc{Flash} compared to a high-resolution, 1D, stationary solution of the hydrodynamic equations at the contact discontinuity. The points in the \textsc{Flash} solution represent temperatures at the centers of individual cells. The radiative cooling rate, \(Q\), was calculated using equation~\ref{eq:Q}.}
    \label{fig:hector_cmp_detail}
\end{figure}

Figure \ref{fig:hector_cmp_detail} shows detailed temperature profiles near the contact discontinuity. In \textsc{Flash}, we used a bubble blown by a~\SI{60}{M_\odot} Milky Way metallicity star expanding into a cloud with \(n_\mathrm{c} = \SI{100}{\centi\meter^{-3}}\). We selected a snapshot at \SI{1.97}{\mega\year}, which is halfway through the star's life. The simulation had a resolution of 8192 cells. The detailed temperature and cooling rate profiles were calculated using a stationary 1D solution (described in Appendix~\ref{app:hector}) of the hydrodynamic equations with boundary conditions from the \ion{H}{ii} and shocked wind regions of the \textsc{Flash} simulation.

The cooling rate is proportional to the density squared and peaks close to the cold medium due to the increasing density. As shown, the area with significant cooling is much smaller than a~single grid cell in our \textsc{Flash} simulations, where numerical mixing creates a cell with the density of the cold medium and a~temperature greater than \(\SI{e4}{\kelvin}\). The entire cell cools rather than just a small part of it. This numerical overcooling acts as an artificial energy sink at the contact discontinuity. By radiating away the thermal energy conducted into the cold shell, it starves the physical mechanism that drives mass evaporation back into the hot interior. As a result, the simulated bubble yields artificially low values for both mass and total energy.

\begin{figure}[htb]
    \centering
    \includegraphics[width=\linewidth]{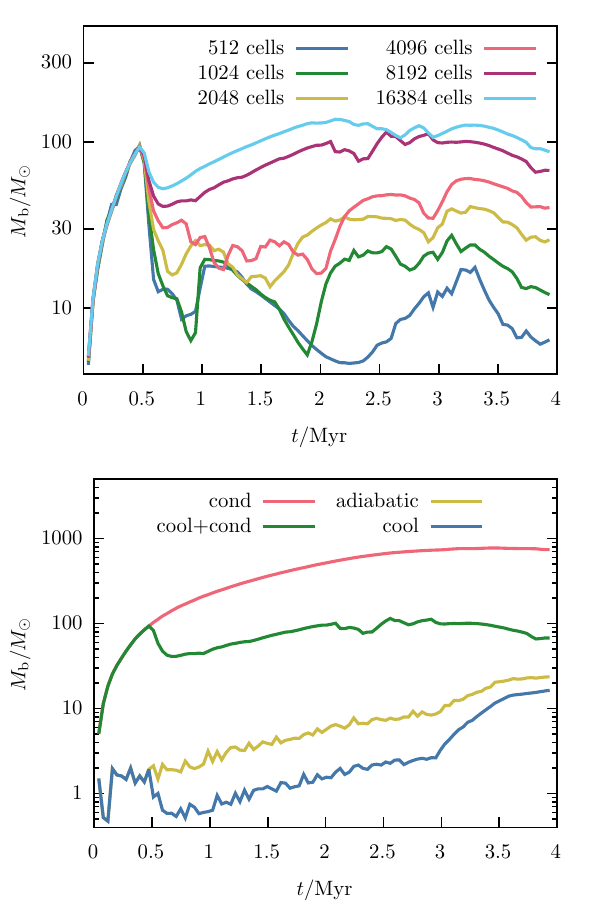}
    \caption{Mass of the hot gas (\(T > \SI{3.5e5}{\kelvin}\), \(p > \SI{0.072}{\kelvin\,\centi\meter^{-3}}\)) inside a~bubble, \(M_\mathrm{b}\), as a function of time. Cooling is turned off for the first \SI{0.5}{\mega\year} in all simulations. The top panel shows a comparison of simulations with thermal conduction and cooling at different resolutions. The mass in the shocked wind region decreases due to overcooling at lower resolutions. The bottom panel shows a~comparison of simulations with 8192 cells with: thermal conduction only, radiative cooling and thermal conduction, no cooling or conduction, and radiative cooling only.}
    \label{fig:convergence_bubble}
\end{figure}

As demonstrated in Figure \ref{fig:convergence_bubble}, there is significantly less mass at lower resolutions. Even at a resolution of 16384 cells, the simulation is not fully converged. Increasing the resolution beyond this point would be computationally expensive. The bottom panel shows that including thermal conduction increases the mass by one to two orders of magnitude, so it cannot be neglected.

In order to reduce overcooling, we impose a limit on the cooling rate. Near the contact discontinuity, we can write the stationary, 1D energy equation as
\begin{equation}
    \frac{\partial}{\partial x} \left ( \rho v \left ( \frac{1}{2} v^2 + \frac{5}{2} c_\mathrm{s}^2 \right ) \right ) + \frac{\partial q_\mathrm{c}}{\partial x} + Q=0\,,
    \label{eq:energyCD}
\end{equation}
where we neglected the work done by gravity. The first term is responsible for the evaporation. At the interface of the contact discontinuity, the evaporation term is positive, the conduction term is negative, and the cooling term is positive. We can put an upper limit on cooling by neglecting the evaporation term
\begin{equation}
    Q \le -\frac{\partial q_\mathrm{c}}{\partial x}\,.
\end{equation}
By substituting \(dx = - \kappa / q_\mathrm{c} dT\), we get
\begin{equation}
    \int_{q_\mathrm{cold}}^{q_\mathrm{hot}} q_\mathrm{c} \, dq_\mathrm{c} = \int_{T_\mathrm{cold}}^{T_\mathrm{hot}} \kappa Q \, dT\,. 
\end{equation}
Far enough at the hot side \(q_\mathrm{hot} \rightarrow 0\). In a stationary state, the energy radiated at the interface cannot be larger than the energy supplied by \(q_\mathrm{cold}\) which can be written as
\begin{equation}
    q_\mathrm{cold} = \sqrt{2 \int_{T_\mathrm{hot}}^{T_\mathrm{cold}} \kappa Q \, dT} = P q_\mathrm{lim}(T_\mathrm{cold}, T_\mathrm{hot}, Z)\,,
\end{equation}
where \(q_\mathrm{lim}\) can be precalculated for each chemical composition, \(T_\mathrm{cold}\), and \(T_\mathrm{hot}\). Here we assume that the cooling layer at the contact discontinuity is roughly isobaric, we substitute \(n \propto P/T\) into the cooling function, allowing pressure to be factored out of the integral. We chose to calculate the tables for \(T_\mathrm{cold} = \SI{e4}{\kelvin}\), but doing so for \(T_\mathrm{cold} = \SI{e2}{\kelvin}\) would yield almost the same results because the contribution of the cooling function to the integral below \SI{e4}{\kelvin} is negligible.

To implement this in \textsc{Flash}, we evaluate a localized search stencil for every cell with a temperature in the range of \SIrange[range-phrase = \text{--}]{e4}{9e5}{\kelvin}. We scan up to 4 cells away in all orthogonal and diagonal directions to determine the maximum neighborhood temperature, \(T_\mathrm{max}\). If \(T_\mathrm{max}\) exceeds double the temperature of the current cell, we look up the corresponding limit, \(q_\mathrm{cold}\), and use it to cap the local cooling rate. Although restricting our search to a 4-cell radius may theoretically underestimate the true \(T_\mathrm{hot}\) and \(q_\mathrm{lim}\), it is physically sufficient. A 4-cell distance easily encompasses the peak of the cooling curve. Because the cooling integral is dominated by this peak, the neglected contributions from more distant, hotter gas are not significant.

\begin{figure}[htb]
    \centering
    \includegraphics[width=\linewidth]{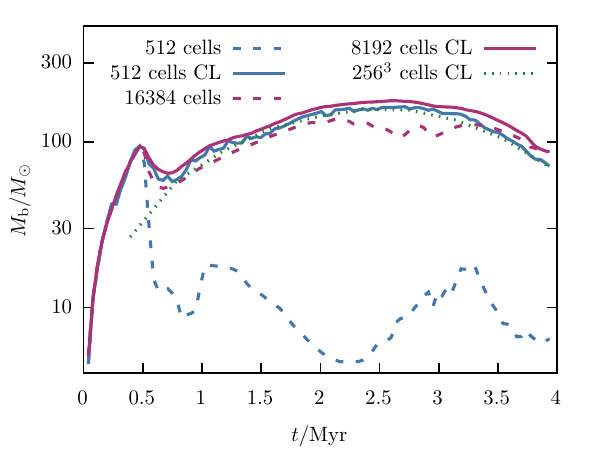}
    \caption{Comparison of simulations with thermal conduction and cooling at different resolutions, showing mass of the hot gas (\(T > \SI{3.5e5}{\kelvin}\), \(p > \SI{0.072}{\kelvin\,\centi\meter^{-3}}\)) inside a~bubble, \(M_\mathrm{b}\), as a function of time. Cooling is turned off for the first \SI{0.5}{\mega\year} in all simulations. Dashed lines are the same as in Figure~\ref{fig:convergence_bubble}, solid lines represent simulations with the cooling limiter, and the dotted line is a 3D simulation with the cooling limiter.}
    \label{fig:converged}
\end{figure}

As demonstrated in Figure \ref{fig:converged}, limiting the cooling brings the results of the simulations at different resolutions very close to the high resolution simulation without a cooling limit. The results are very similar, even for a~very low-resolution 3D simulation. The small differences may be due to variations in \(T_\mathrm{max}\), because at lower resolutions, we can see farther away into the hot bubble within the 4-cell search radius. Furthermore, since the imposed cooling limit is only an upper limit, there may still be some overcooling, albeit by orders of magnitude less.

\section{1D and 3D accretion comparison}
\label{app:accretion_comparison}

\begin{figure}[htb]
    \centering
    \includegraphics[width=\linewidth]{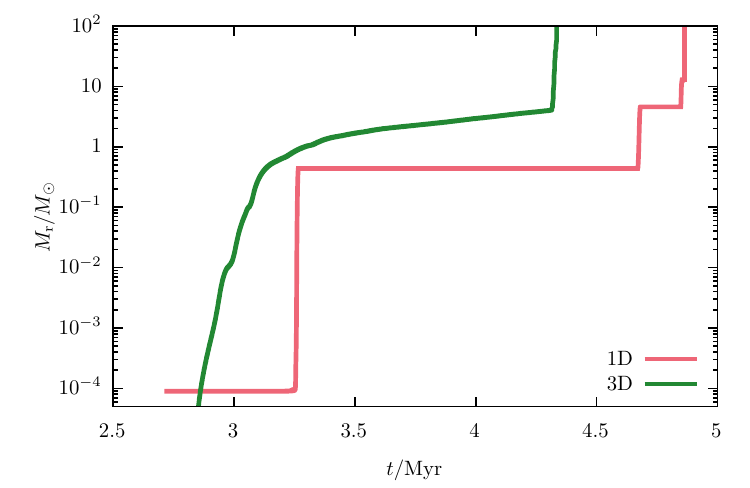}
    \caption{Re-accreted mass, \(M_\mathrm{r}\), over time for 1D and 3D models. This tracks the post-collapse evolution of a \SI{100}{M_\odot} progenitor bubble (\(n_\mathrm{c} = \SI{500}{\centi\meter^{-3}}\), \(E_\mathrm{grav} = \SI{e50}{\erg}\)).}
    \label{fig:accretion_comp}
\end{figure}

Figure \ref{fig:accretion_comp} compares the re-accreted mass, \(M_\mathrm{r}\), over time for 1D and 3D simulations. Relaxing forced spherical symmetry in 3D yields a more gradual accretion process than the discrete episodes seen in 1D. Accretion happens slightly earlier in 3D due to moderate overcooling and faster bubble collapse. Both tracks end with a sharp increase as the massive outer shell collapses. Because modeling this final collapse is outside of our current framework, we evaluated final re-accreted masses just prior to this.

\section{1D parameter space maps}
\label{app:pspace}

\begin{figure}[htb]
    \centering
    \(M_* = \SI{60}{M_\odot}, Z_* = \SI{0.016}{Z_\odot}\)
    \includegraphics[width=\linewidth]{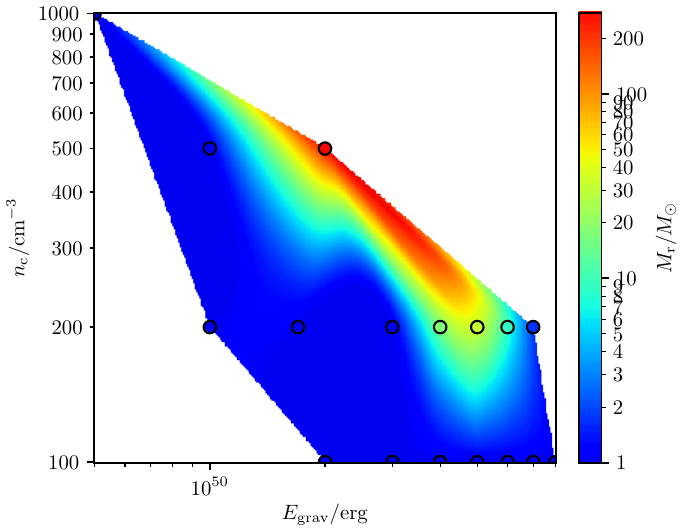}
    \(M_* = \SI{150}{M_\odot}, Z_* = \SI{0.016}{Z_\odot}\)
    \includegraphics[width=\linewidth]{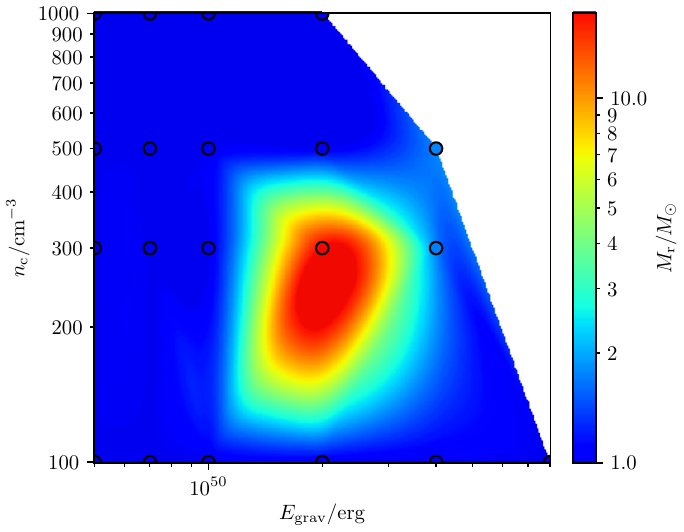}
    \caption{Mass re-accreted onto a BH, \(M_\mathrm{r}\), formed by the direct collapse of a star as a function of the initial gravitational energy of the cloud, \(E_\mathrm{grav}\), and the initial particle density of the cloud, \(n_\mathrm{c}\). Black circles show individual simulations. Color maps and contours were calculated using cubic interpolation in logarithmic space from the individual simulations.}
    \label{fig:nosn_reacc_app}
\end{figure}

\begin{figure*}
    \centering
    \begin{minipage}{0.33\textwidth}
        \centering
        \(M_* = \SI{60}{M_\odot}, Z_* = \SI{0.66}{Z_\odot}\)
        \includegraphics[width=\textwidth]{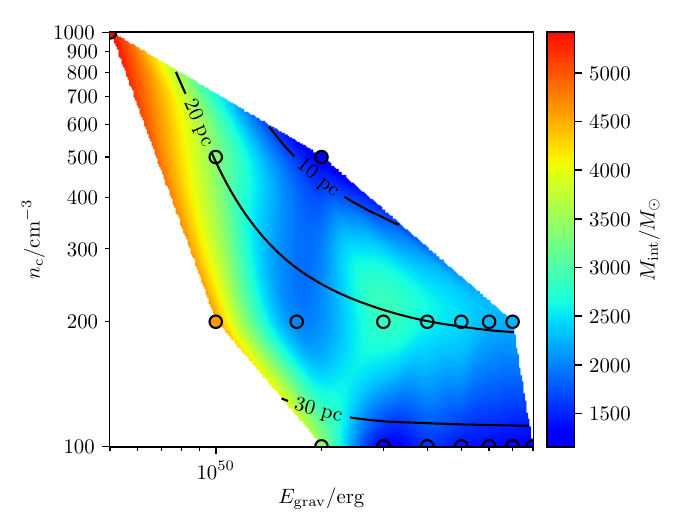}
    \end{minipage}%
    \hfill%
    \begin{minipage}{0.33\textwidth}
        \centering
        \(M_* = \SI{120}{M_\odot}, Z_* = \SI{0.66}{Z_\odot}\)
        \includegraphics[width=\textwidth]{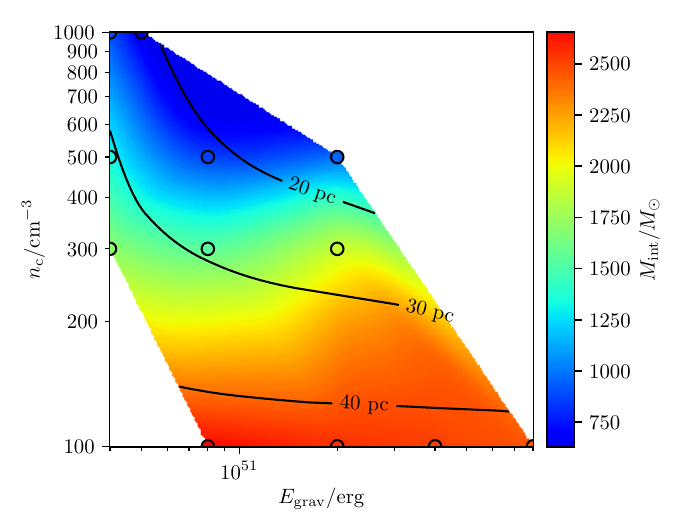}
    \end{minipage}%
    \hfill%
    \begin{minipage}{0.33\textwidth}
        \centering
        \(M_* = \SI{250}{M_\odot}, Z_* = \SI{0.66}{Z_\odot}\)
        \includegraphics[width=\textwidth]{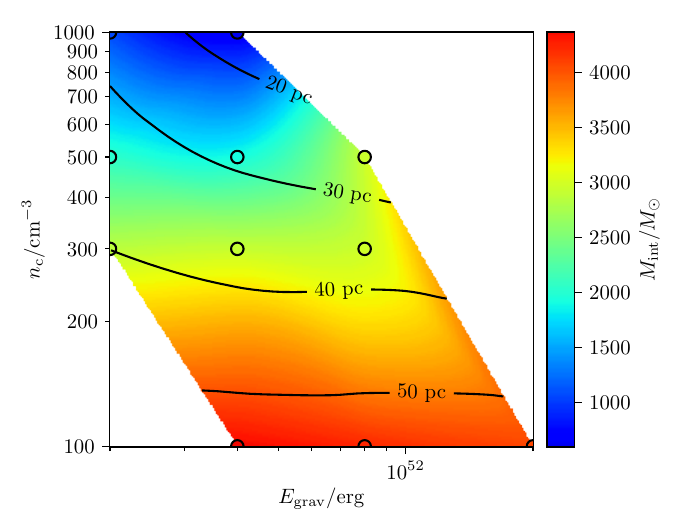}
    \end{minipage}
    \begin{minipage}{0.33\textwidth}
        \centering
        \includegraphics[width=\textwidth]{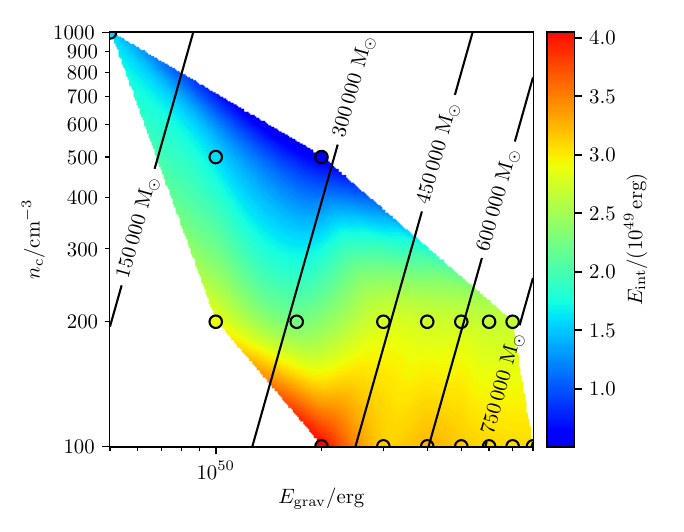}
    \end{minipage}%
    \hfill%
    \begin{minipage}{0.33\textwidth}
        \centering
        \includegraphics[width=\textwidth]{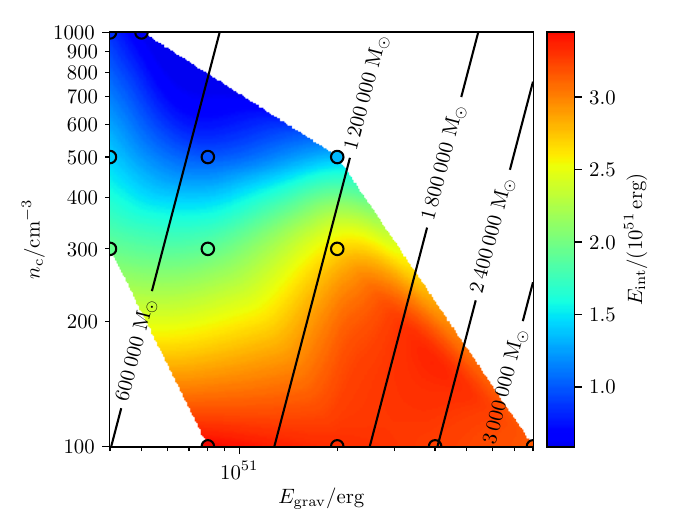}
    \end{minipage}%
    \hfill%
    \begin{minipage}{0.33\textwidth}
        \centering
        \includegraphics[width=\textwidth]{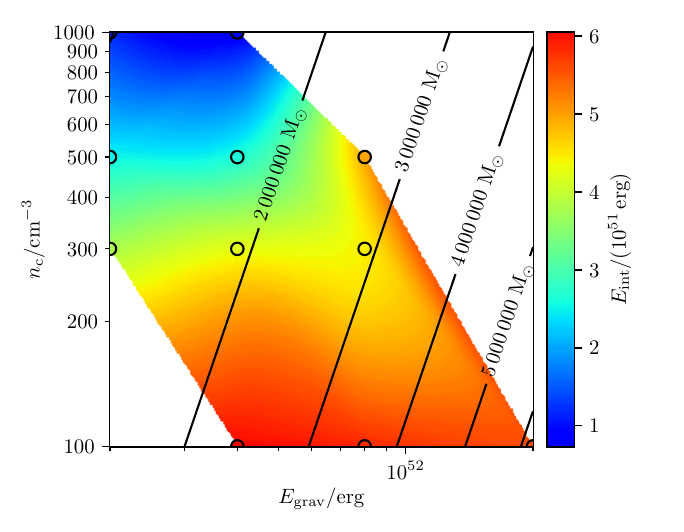}
    \end{minipage}
    \begin{minipage}{0.33\textwidth}
        \centering
        \(M_* = \SI{45}{M_\odot}, Z_* = \SI{0.016}{Z_\odot}\)
        \includegraphics[width=\textwidth]{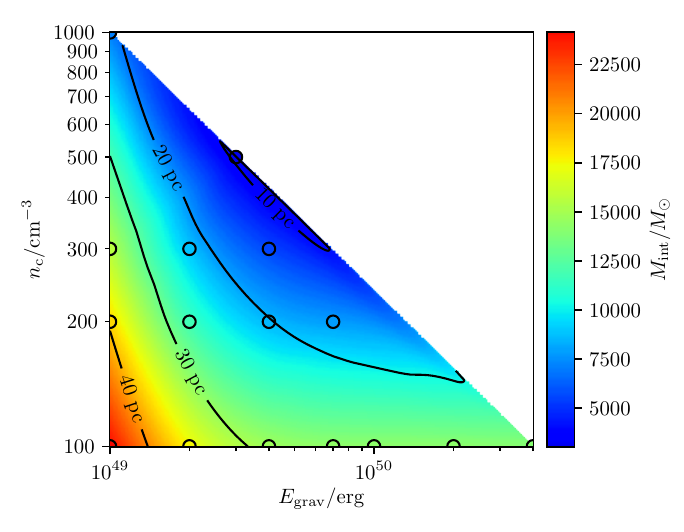}
    \end{minipage}%
    \hfill%
    \begin{minipage}{0.33\textwidth}
        \centering
        \(M_* = \SI{77}{M_\odot}, Z_* = \SI{0.016}{Z_\odot}\)
        \includegraphics[width=\textwidth]{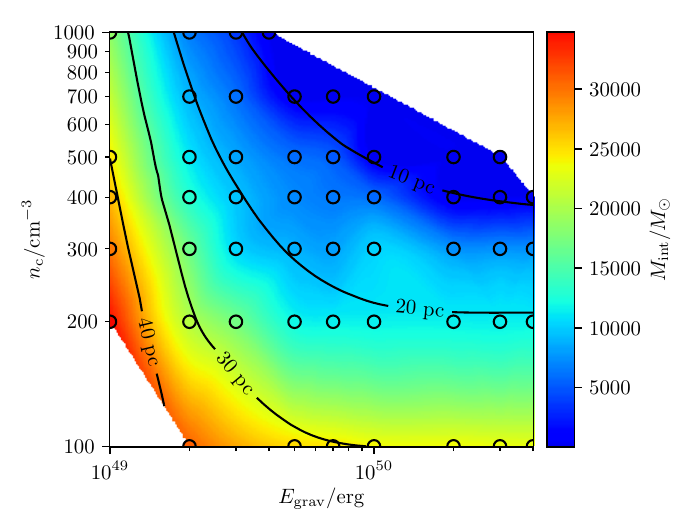}
    \end{minipage}%
    \hfill%
    \begin{minipage}{0.33\textwidth}
        \centering
        \(M_* = \SI{150}{M_\odot}, Z_* = \SI{0.016}{Z_\odot}\)
        \includegraphics[width=\textwidth]{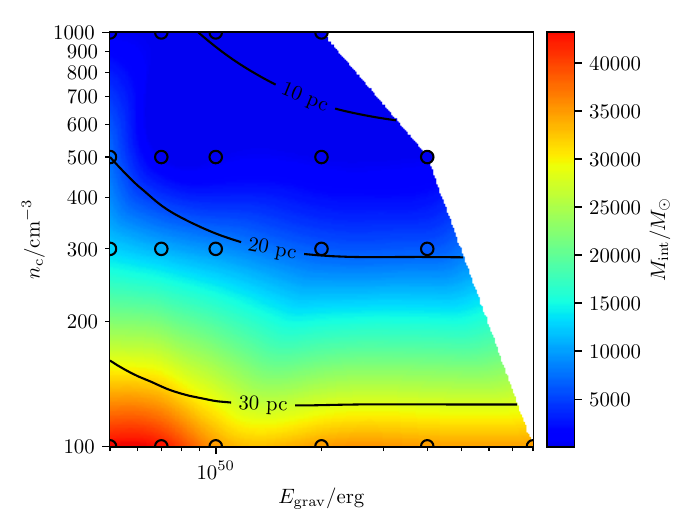}
    \end{minipage}
    \begin{minipage}{0.33\textwidth}
        \centering
        \includegraphics[width=\textwidth]{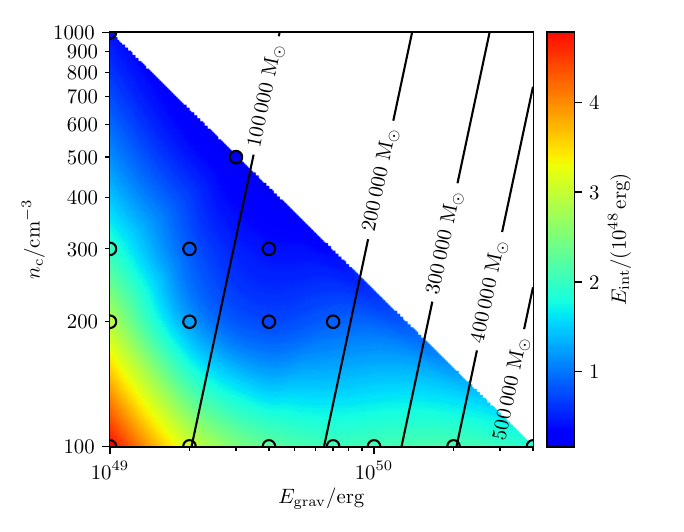}
    \end{minipage}%
    \hfill%
    \begin{minipage}{0.33\textwidth}
        \centering
        \includegraphics[width=\textwidth]{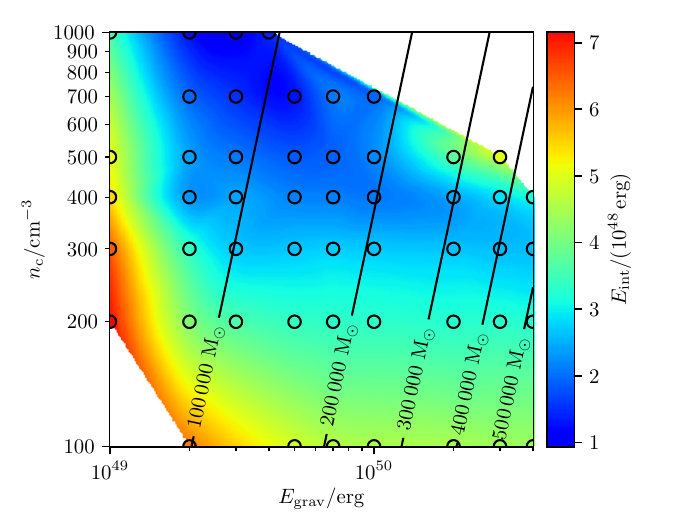}
    \end{minipage}%
    \hfill%
    \begin{minipage}{0.33\textwidth}
        \centering
        \includegraphics[width=\textwidth]{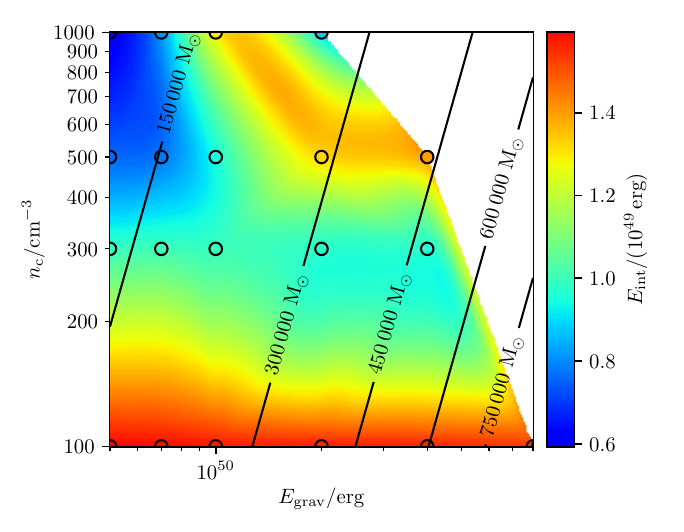}
    \end{minipage}
    \caption{Properties of the bubbles \SI{100}{\kilo\year} after the star collapses into a~BH as a function of the initial gravitational energy of the cloud, \(E_\mathrm{grav}\), and the initial particle density of the cloud, \(n_\mathrm{c}\). The first and third rows show the mass of the gas under the dense shell, \(M_\mathrm{int}\). The second and last rows show the internal energy of the gas inside the bubble excluding the shell, \(E_\mathrm{int}\), at the same time as the top panels. Contours show the radius of the shell and the initial mass of the cloud. Black circles show individual simulations. Color maps and contours were calculated using cubic interpolation from the individual simulations.}
    \label{fig:nosn_maps_app}
\end{figure*}

\begin{figure*}
    \centering
    \begin{minipage}{0.33\textwidth}
        \centering
        \(M_* = \SI{60}{M_\odot}, Z_* = \SI{0.66}{Z_\odot}\)
        \includegraphics[width=\textwidth]{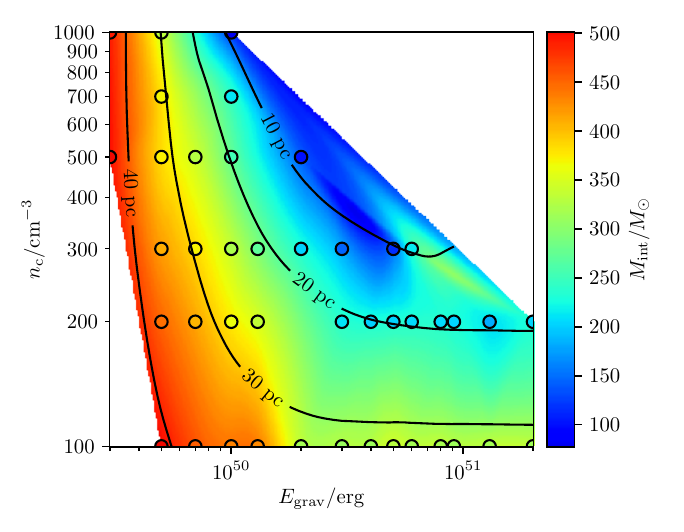}
    \end{minipage}%
    \hfill%
    \begin{minipage}{0.33\textwidth}
        \centering
        \(M_* = \SI{120}{M_\odot}, Z_* = \SI{0.66}{Z_\odot}\)
        \includegraphics[width=\textwidth]{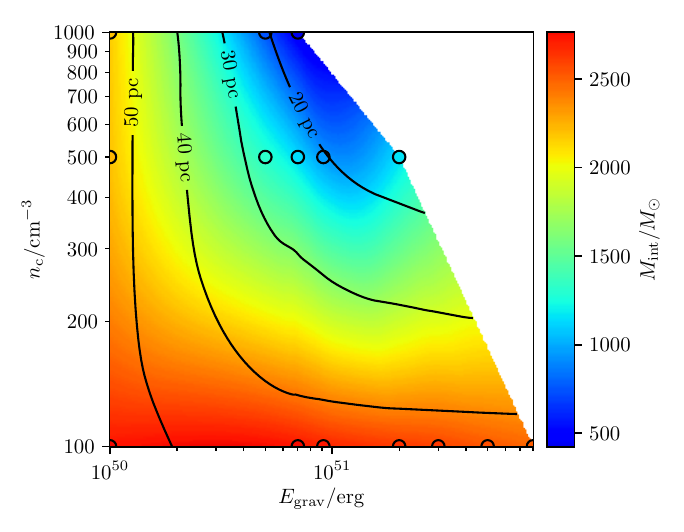}
    \end{minipage}%
    \hfill%
    \begin{minipage}{0.33\textwidth}
        \centering
        \(M_* = \SI{250}{M_\odot}, Z_* = \SI{0.66}{Z_\odot}\)
        \includegraphics[width=\textwidth]{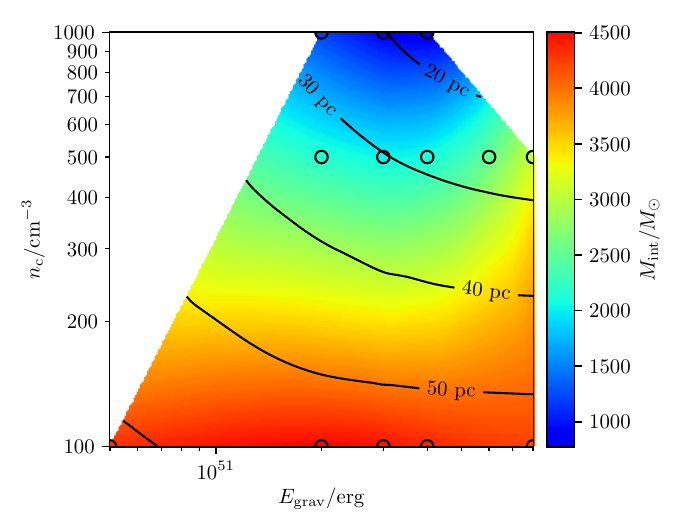}
    \end{minipage}
    \begin{minipage}{0.33\textwidth}
        \centering
        \includegraphics[width=\textwidth]{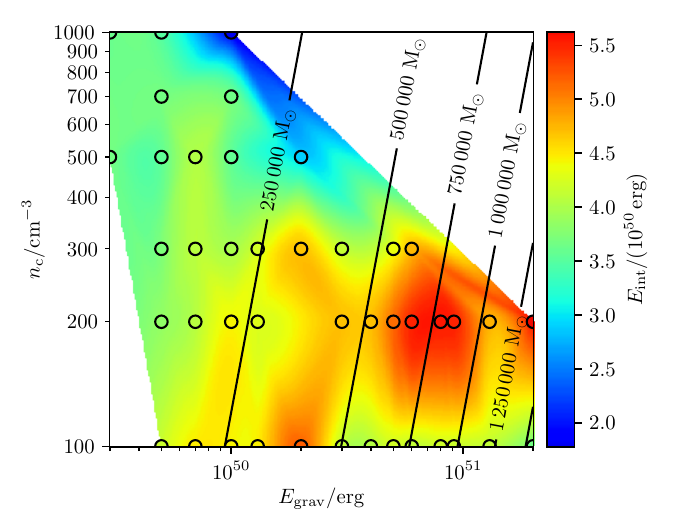}
    \end{minipage}%
    \hfill%
    \begin{minipage}{0.33\textwidth}
        \centering
        \includegraphics[width=\textwidth]{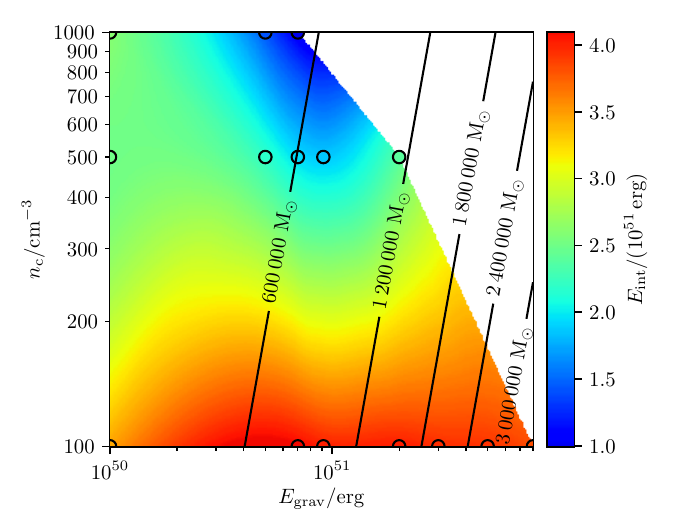}
    \end{minipage}%
    \hfill%
    \begin{minipage}{0.33\textwidth}
        \centering
        \includegraphics[width=\textwidth]{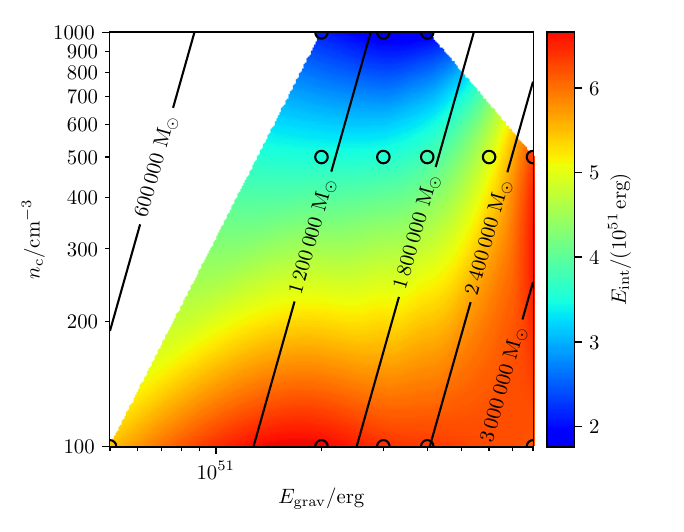}
    \end{minipage}
    \begin{minipage}{0.33\textwidth}
        \centering
        \(M_* = \SI{45}{M_\odot}, Z_* = \SI{0.016}{Z_\odot}\)
        \includegraphics[width=\textwidth]{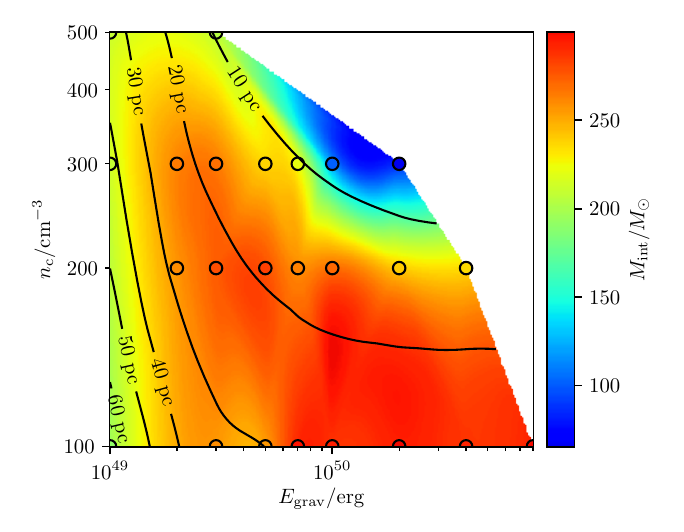}
    \end{minipage}%
    \hfill%
    \begin{minipage}{0.33\textwidth}
        \centering
        \(M_* = \SI{77}{M_\odot}, Z_* = \SI{0.016}{Z_\odot}\)
        \includegraphics[width=\textwidth]{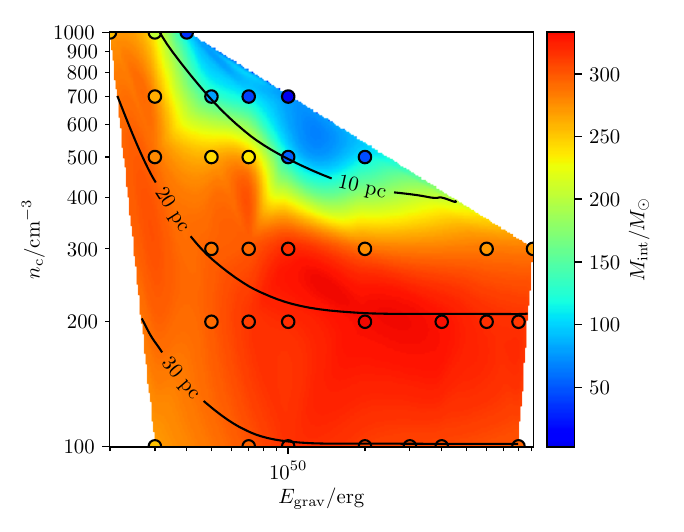}
    \end{minipage}%
    \hfill%
    \begin{minipage}{0.33\textwidth}
        \centering
        \(M_* = \SI{150}{M_\odot}, Z_* = \SI{0.016}{Z_\odot}\)
        \includegraphics[width=\textwidth]{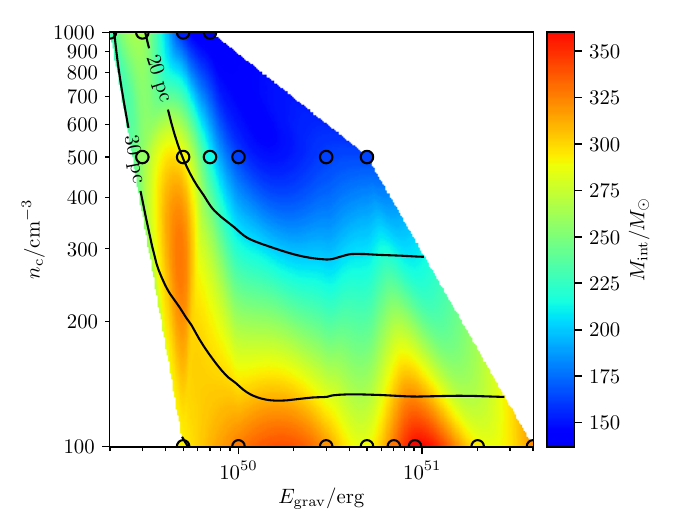}
    \end{minipage}
    \begin{minipage}{0.33\textwidth}
        \centering
        \includegraphics[width=\textwidth]{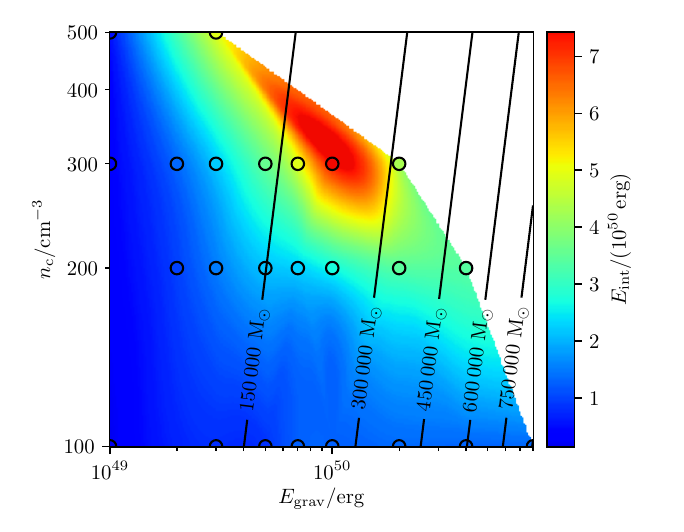}
    \end{minipage}%
    \hfill%
    \begin{minipage}{0.33\textwidth}
        \centering
        \includegraphics[width=\textwidth]{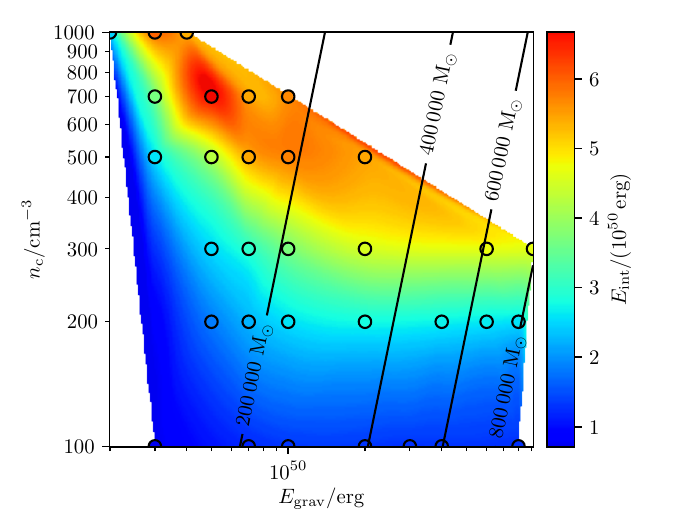}
    \end{minipage}%
    \hfill%
    \begin{minipage}{0.33\textwidth}
        \centering
        \includegraphics[width=\textwidth]{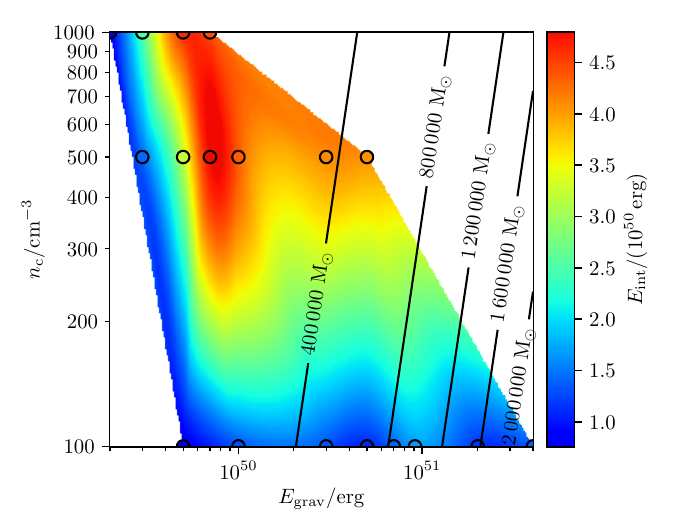}
    \end{minipage}
    \caption{Properties of the bubbles when the SN ejecta reflects from the dense shell as a function of the initial gravitational energy of the cloud, \(E_\mathrm{grav}\), and the initial particle density of the cloud, \(n_\mathrm{c}\). The first and third rows show the mass of the gas under the dense shell, \(M_\mathrm{int}\). The second and last rows show the internal energy of the gas inside the bubble excluding the shell, \(E_\mathrm{int}\). The contours show the radius of the shell and the initial mass of the cloud. Black circles show individual simulations. Color maps and contours were calculated using cubic interpolation from the results of individual simulations.}
    \label{fig:sn_maps_app}
\end{figure*}

Figure \ref{fig:nosn_reacc_app} shows the mass re-accreted onto a BH formed by the direct collapse of a star.
Figure \ref{fig:nosn_maps_app} shows the properties of the bubbles \SI{100}{\kilo\year} after the star collapsed into a~BH.
Figure \ref{fig:sn_maps_app} shows the properties of the bubbles at the time when the SN ejecta reaches the cold, dense shell.

\newpage

\section{Stationary 1D solution at the contact discontinuity}
\label{app:hector}

To find a stationary 1D solution at the contact discontinuity, we rewrite equations \ref{eq:f1}--\ref{eq:E} in 1D, without time derivatives and the source terms as
\begin{equation}
    \frac{\partial}{\partial x}(\rho v) = 0\,,
    \label{eq:cont}
\end{equation}
\begin{equation}
    \rho v \frac{dv}{dx} = -\frac{dp}{dx}\,,
    \label{eq:motion}
\end{equation}
\begin{equation}
    \frac{\partial}{\partial x} \left ( \rho v \left ( \frac{1}{2} v^2 + \frac{5}{2} c_\mathrm{s}^2 \right ) \right ) + \frac{\partial q_\mathrm{c}}{\partial x} + n^2 \Lambda(T, Z)=0\,,
    \label{eq:energy}
\end{equation}
where the pressure, \(p\), is given by Equation \ref{eq:eos}, the speed of sound, \(c_\mathrm{s}\), is given by
\begin{equation}
    c_\mathrm{s}^2 = \frac{\gamma k_\mathrm{B} T}{\mu m_\mathrm{H}}\,,
    \label{eq:sound}
\end{equation}
and the thermal conduction flux is
\begin{equation}
    q_\mathrm{c} = -K_\mathrm{eff} T^{\frac{5}{2}} \frac{dT}{dx}\,,
    \label{eq:conductivity}
\end{equation}
where \(K_\mathrm{eff}\) is calculated using Equation \ref{eq:keff}. Equations \ref{eq:cont}--\ref{eq:conductivity} can be expressed as a system of four first-order differential equations we solve numerically using the explicit 8th-order Runge--Kutta method (DOP853 scheme; \citet{hairer_solving_1993}).

\section{Sensitivity to \(\mu\)}
\label{app:mu}

\begin{figure*}[p]
    \centering
    \begin{minipage}{0.33\textwidth}
        \centering
        \includegraphics[width=\textwidth]{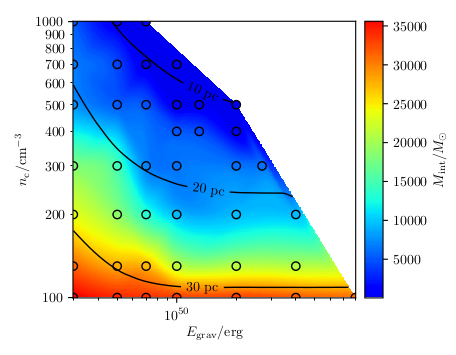}
    \end{minipage}%
    \hfill%
    \begin{minipage}{0.33\textwidth}
        \centering
        \includegraphics[width=\textwidth]{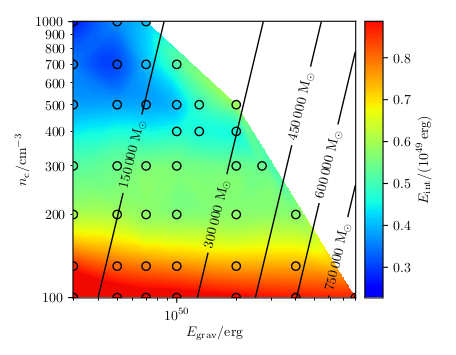}
    \end{minipage}%
    \hfill%
    \begin{minipage}{0.33\textwidth}
        \centering
        \includegraphics[width=\textwidth]{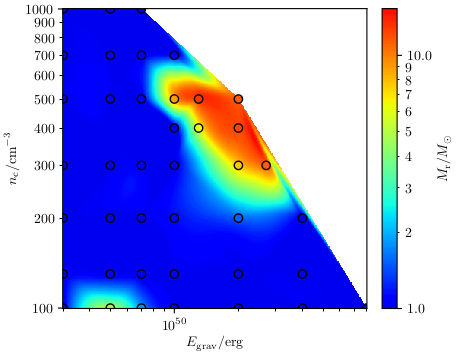}
    \end{minipage}
    \begin{minipage}{0.33\textwidth}
        \centering
        \includegraphics[width=\textwidth]{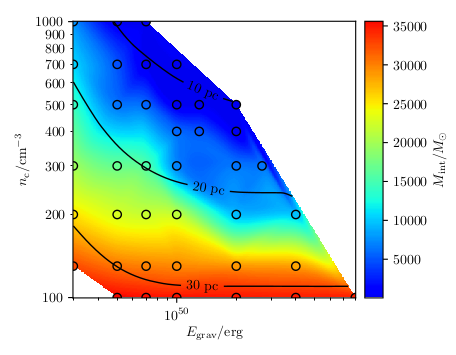}
    \end{minipage}%
    \hfill%
    \begin{minipage}{0.33\textwidth}
        \centering
        \includegraphics[width=\textwidth]{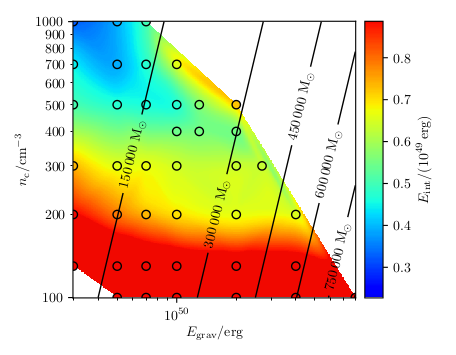}
    \end{minipage}%
    \hfill%
    \begin{minipage}{0.33\textwidth}
        \centering
        \includegraphics[width=\textwidth]{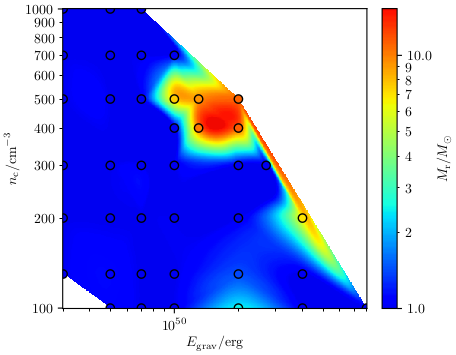}
    \end{minipage}
    \caption{1D parameter space study results for the direct collapse case of a \SI{100}{M_\odot} star with the metallicity of I Zwicky 18. The first row shows simulations without molecular gas (neutral atomic and ionized only), while the second row shows simulations where all gas is ionized. The first column shows the mass inside the bubble, the second column the internal energy, and the last column the total accreted mass.}
    \label{fig:pspace_noSN}
\end{figure*}

\begin{figure*}[p]
    \centering
    \begin{minipage}{0.4\textwidth}
        \centering
        \includegraphics[width=\textwidth]{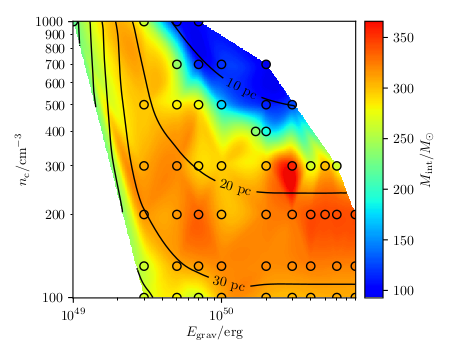}
    \end{minipage}%
    \hfill%
    \begin{minipage}{0.4\textwidth}
        \centering
        \includegraphics[width=\textwidth]{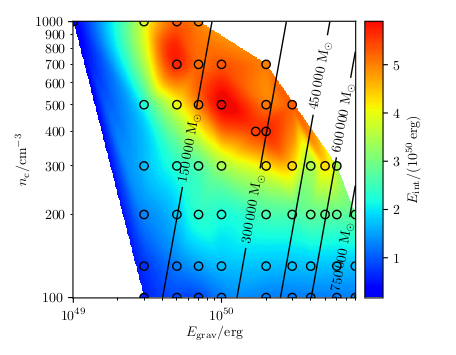}
    \end{minipage}
    \begin{minipage}{0.4\textwidth}
        \centering
        \includegraphics[width=\textwidth]{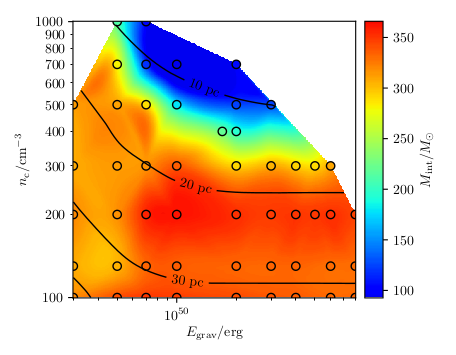}
    \end{minipage}%
    \hfill%
    \begin{minipage}{0.4\textwidth}
        \centering
        \includegraphics[width=\textwidth]{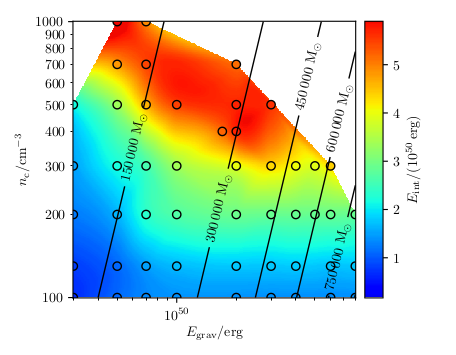}
    \end{minipage}
    \caption{1D parameter space study results for the SN case for a \SI{100}{M_\odot} star with the metallicity of I Zwicky 18. The first row shows simulations without molecular gas (neutral atomic and ionized only), while the second row shows simulations where all gas is ionized. The first column shows the mass inside the bubble and the second column the internal energy.}
    \label{fig:pspace_SN}
\end{figure*}

As stated in Section~\ref{sec:model}, we make simplifications when calculating the chemical state of the gas and calculate \(\mu\) only from the temperature of the gas. To verify that our results are not sensitive to the exact atomic state, we recalculated the 1D parameter space for the \SI{100}{M_\odot} star with the metallicity of I Zwicky 18 for the following cases:
\begin{itemize}
    \item There is no molecular gas. This case is shown in the first row of Figures \ref{fig:pspace_noSN} and \ref{fig:pspace_SN}.
    \item Gas is allowed to be only ionized. This case is shown in the second row of Figures \ref{fig:pspace_noSN} and \ref{fig:pspace_SN}.
\end{itemize}

By not allowing the gas to be molecular (or neutral in the second case), we put an upper limit on the value of \(\mu\). This increases the pressure in the dense gas. Consequently, the region with significant re-accretion slightly shrinks; however, the overall qualitative result remains the same, including regions with around \SI{10}{M_\odot} of re-accretion. The properties of the bubbles remain very close to the baseline case. The biggest difference can be seen in the internal energy of the direct collapse case with only ionized gas. The higher pressure inside the \ion{H}{ii} region, which does not become neutral after the star collapses, reduces the density and thus cooling.

We conclude that the outcome of our simulations is not very sensitive to the exact value of \(\mu\), and our simplifications are not significantly changing the results.

\end{appendix}

\end{document}